\documentclass[alpha-refs]{nbdt-article}
\usepackage[american]{babel}
\usepackage{csquotes}
\usepackage[style=apa,natbib=true,backend=biber]{biblatex}

\usepackage{xassoccnt}
\newcounter{totalequations}
\DeclareAssociatedCounters{equation}{totalequations}%

\let\theOldHequation\theHequation
\renewcommand{\theHequation}{\theOldHequation::\number\value{totalequations}}

\pdfoutput=1
\usepackage{siunitx}
\usepackage{upgreek}
\usepackage{amsmath}
\usepackage{bm}
\usepackage{doi}

\usepackage{xcolor}
\hypersetup{
    colorlinks,
    linkcolor={red!50!black},
    citecolor={blue!50!black},
    urlcolor={blue!80!black}
}

\papertype{Original Article}

\title{Probabilistic representations as building blocks for higher-level vision}

\abbrevs{FDL, feature distribution learning; MEO, mean expected orientation.}

\author[1]{Andrey Chetverikov}
\author[2]{Árni Kristjánsson}

\affil[1]{Donders Institute for Brain, Cognition and Behavior, Radboud University, The Netherlands}
\affil[2]{Icelandic Vision Lab, Faculty of Psychology, University of Iceland, Iceland}

\corraddress{Andrey Chetverikov, Donders Institute for Brain, Cognition and Behavior, Radboud University, The Netherlands; Árni Kristjánsson, Faculty of Psychology, University of Iceland, Iceland}
\corremail{a.chetverikov@donders.ru.nl; ak@hi.is}

\fundinginfo{Supported by a grant from the Icelandic Research Fund (IRF \#173947-052)
and a grant from the Russian Foundation for Basic Research (RFBR,
\#15-36-01358, 2015-2017). AC was supported by the Radboud Excellence Initiative.}
\runningauthor{Chetverikov \& Kristjánsson}

\begin{document}

\maketitle

\begin{abstract}
Current theories of perception suggest that the brain represents features of the world as probability distributions, but can such uncertain foundations provide the basis for everyday vision? Perceiving objects and scenes requires knowing not just how features (e.g., colors) are distributed but also where they are and which other features they are combined with. Using a Bayesian computational model, we recovered probabilistic representations used by human observers to search for odd stimuli among distractors. Importantly, we found that the brain integrates information between feature dimensions and spatial locations, leading to more precise representations compared to when information integration is not possible. We also uncovered representational asymmetries and biases, showing their spatial organization and explain how this structure argues against ``summary statistics'' accounts of visual representations. Our results confirm that probabilistically encoded visual features are bound with other features and to particular locations, providing a powerful demonstration of how probabilistic representations can be a foundation for higher-level vision.

\keywords{probabilistic perception, binding problem, ensemble perception, summary statistics, visual search}
\end{abstract}

\section{Introduction}

How the brain represents the visual world is a long-standing question in cognitive science. One captivating idea is that the brain builds statistical models that describe probability distributions of visual features in the environment \citep{tanrikulu2021What, rao2002Probabilistic, zemel1998Probabilistic, fiser2010Statistically, lange2020Bayesian, knill2004Bayesian, pouget2000Information}. By combining information about different features and their locations, the brain can then form representations of objects and scenes. Indeed, the idea that the brain represents feature distributions matches our conscious visual experience well. Most objects, such as the apple in Figure \ref{fig:general_approach}A, contain a multitude of feature values that can be quantified as a probability distribution, and we are seemingly aware of these feature constellations. Surprisingly, most studies of probabilistic representations do not test how such constellations are represented, assuming instead that a stimulus is described by a single value, such as the orientation of a Gabor patch in vision studies or the hue of an item in working memory experiments and that the only uncertainty comes from the sensory noise. While this unrealistic assumption was noted a while ago \citep{zemel1998Probabilistic}, it is still prevalent, leaving open the possibility that the results can be explained with alternative models without assuming detailed representations of probability distributions \citep{block2018If, rahnev2017case}.

Here, we aim to close this gap and ask 1) if the visual system is capable of quickly forming precise representations of heterogeneous stimuli, representations that reflect the probability distribution of their features and 2) if such representations can be bound to other features or to spatial locations thereby serving as building blocks for upstream object and scene processing.

What precisely do we mean by probabilistic perceptual representations? We assume that the brain operates with probabilistic representations if any of the internal variables used in the perceptual decision-making is represented probabilistically, that is, allowing for uncertainty in their values (similar to, e.g., \citealp{koblinger2021Representations}). Often, the concept of probabilistic representations is embedded in the context of Bayesian models. Bayesian perceptual models assume that observers are presented with a stimulus \emph{s} that generates sensory observations \emph{x} with a certain probability \emph{p(x\textbar s)}. The observer knows the parameters of this \emph{generative model}, that is, \emph{p(x\textbar s)}, and can inverse it to compute \emph{p(s\textbar x)}, the probability that a distal stimulus \emph{s} has a certain value given a sensory observation \emph{x}. Importantly, this approach focuses on how a stimulus is inferred from sensory observations. This inferred probability \emph{p(s\textbar x)} is associated with the probabilistic representation. This is intuitively agreeable as long as the focus is on a simple single stimulus but becomes murky when stimuli are heterogenous like the ones shown in Figure \ref{fig:general_approach} (and in reality, homogeneous stimuli do not exist) as it is not clear what an observer infers or should infer in this case.

We approach the problem of probabilistic representations differently, asking instead whether an observer can represent a probability that a stimulus (e.g., an apple or a set of lines in Figure \ref{fig:general_approach}) could have a feature with a certain value (e.g., the red color on an apple or a line with a certain orientation). For example, are apples likely to contain red? In Bayesian terms, this means extending the model outlined above to a stimulus-feature-observation hierarchy and asking whether observers can represent probability distributions within this hierarchy, specifically, whether a probability distribution of features given a heterogeneous stimulus is represented within an observers' generative model.

\subsection{Probabilistic representations of heterogenous stimuli}

\begin{figure}[bt]
\centering
\includegraphics[width=\textwidth,height=\textheight,keepaspectratio]{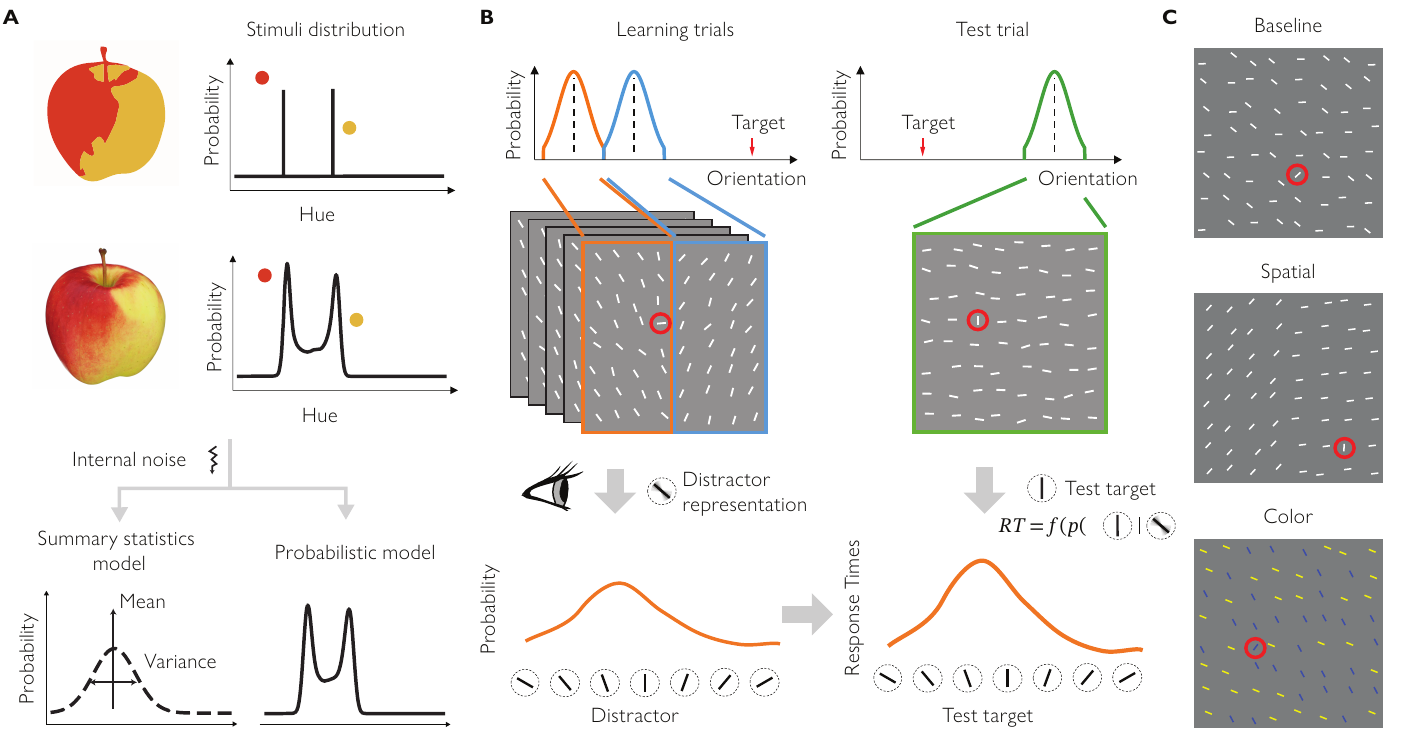}
\caption{General approach and methods. \textbf{A}: A typical stimulus used to study probabilistic perception involves an impoverished version of the environment, similar to a sketch of an apple (top-left). The hues of this stimulus can be quantified as a discrete probability distribution with only a few probable values (top-right). In contrast, real objects have a multitude of feature values corresponding to a complex-shaped probability distribution (middle). An accurate probabilistic model would maintain the important details of the distribution as much as internal noise permits, while a summary statistics model suggests that probabilities are represented as a combination of simple parameters, such as mean and variance (bottom). \textbf{B}: In our experiments, in each miniblock (consisting of a sequence of learning trials and 1 or 2 test trials) observers searched for an odd-one-out line among distractors. On learning trials (upper-left), distractors were drawn from two distributions that were either mixed together or separated by location or color with one example of the spatial separation shown here. We assumed that observers would form a distractor representation by learning which distractors are more probable as shown in previous studies (bottom-left). On test trials (upper-right), we varied the similarity between the target and previously learned distractors. We then measured response times assuming that they should be monotonically related to the probability of a given target being a distractor based on a simplified ideal observer model (bottom-right). \textbf{C}: Example stimuli used on learning trials in Experiment 1.}
\label{fig:general_approach}
\end{figure}

How can the brain represent heterogeneous stimuli, that is, stimuli that have more than one feature value? The visual system may track each feature value at each location to form a precise representation isomorphic to the stimulus. However, this would be extremely costly in terms of computational resources and unnecessary or even misleading for action because specific feature values can vary from one moment to another because of changes in viewpoint, lighting, etc. \citep{kristjansson2022Priming}. Another possibility, explored in the ``summary statistics''
\footnote{Note that this is different from image-computable summary statistics approaches based on the statistics of the outputs from multi-level image processing filters \citep{balas2009summarystatistic, freeman2011Metamers, portilla2000Parametric}. While these are related, the statistics in the ensemble perception literature are conceptualized in a more abstract way, more consistent with the type of questions we are interested in here. Yet, even in the image-computable statistics literature it has been demonstrated that images identical in a model statistical space might be still distinguishable by the observers, suggesting that the image-computable summary statistics do not fully match human perception \citep{wallis2016Testing}} 
or ``ensemble perception'' literature \citep{ariely2001Seeing, cohen2016What, haberman2012Ensemble, rahnev2017case, treisman2006How} is that only a few values, for example, the mean and the variance are represented. Note that the concept of probabilistic representations is rarely used in this field, because an observer can compute summary statistics without using probability distributions. For example, an average of the features can be computed arithmetically from sensory observations. However, such representations are functionally equivalent to a simplified probability distribution (Figure \ref{fig:general_approach}A). But we believe that such simplified representations are also unlikely because multiple stimuli can have the same summary statistics while being quite different from each other. More realistically, the brain could compromise by approximating feature distributions in the responses of neuronal populations that capture important aspects of stimuli without being too detailed (Figure \ref{fig:general_approach}A).

Previous studies have indeed shown that the visual system encodes the approximate distribution of visual features and uses them in perceptual decision-making \citep{girshick2011Cardinal, series2013Learning}. However, most of the findings are confined to relatively long-term learning of environmental statistics. If feature probability distributions are to be useful for everyday visual tasks, such as object recognition or scene segmentation, the brain needs to learn feature distributions quickly and effortlessly. Importantly, we have recently provided evidence that such rapid learning may occur in simple cases by studying how human observers learn to ignore distracting stimuli while searching the visual scene \citep{chetverikov2016Building, chetverikov2019Feature, chetverikov2020Probabilistic, chetverikov2017Representing, hansmann-Roth2019Representing, tanrikulu2021Testing}. The basic idea with this \emph{Feature Distribution Learning} paradigm is to use role-reversal effects upon response times when targets and distractors change their roles between visual search trials, to reveal observers expectations about upcoming search displays \citep{chetverikov2019Feature}. Priming in visual search is a well-known phenomenon: search is faster when features of targets or distractors repeat from trial-to-trial even when observers do not have to rely on previous trials, that is, when a target is defined as an odd-one-out \citep{kristjansson2008Priming, kristjansson2022Priming, kristjansson2010Where, lamy2008Priming, maljkovic1994Priming}. And when the targets and distractor switch features (``role reversal''), the search is slower. In our previous experiments, observers were asked to find an odd-one-out item in a search array where, importantly, distractor features (colors or orientations) are randomly drawn from a given probability distribution for several trials rather than having constant features. A test trial (introducing the role reversal) is then presented with a target of varying similarity to previously learned distractors. We found that response times as a function of this similarity parameter followed the shape of the previously learned probability distribution, whether it was Gaussian, uniform, skewed, or even bimodal. That is, the search was slowed proportionally to how unexpected the target was, based on previously learned environmental statistics. This shows that representations of the shape of feature probability distributions in the visual input (similar to scene statistics \citep{oliva2001Modeling, rosenholtz2016Capabilities}) are not limited to long-term learning, but can occur rapidly.

This previous work was, however, limited to simple scenarios with a single feature distribution present, while real environments contain multiple objects (that contain multiple features) and scene parts with various different features. Furthermore, knowledge about statistics of a given feature (e.g., orientation) in isolation is not very useful. Observers need to know \emph{where} in the external world a given feature distribution is and which other features should be bound with it (related to the ``binding'' problem, \citep{treisman1996binding}) to recognize objects or segment scenes. Notably, such binding to spatiotopic locations and to other features does not necessarily require any additional neural machinery, because information about feature distributions can be readily encoded in neural population responses \citep{pouget2000Information, sahani2003Doubly, vertes2018Flexible, zemel1998Probabilistic}. Evidence for such effortless integration of probabilistic visual inputs is, however, still lacking.

Ensemble averaging studies testing how observers estimate probabilistic properties of several sets of stimuli provide some initial support for the hypothesis that probabilistic information can be bound to locations or other features. It is well known that observers can estimate the average of a perceptual ensemble, such as the mean orientation of a set of lines \citep{alvarez2011Representing, haberman2012Ensemble, whitney2018Ensemble}. Notably, they can estimate properties of subsets grouped by location or by other features although this causes performance detriments \citep{attarha2014Summary, attarha2015capacity, attarha2015perceptual, chong2005Statistical, oriet2013Size, utochkin2017numerosity}. This means that at least a summary representation, functionally equivalent to a simplified probabilistic representation based on mean and variance, can be bound to a location. If, for example, a mean can be computed only for a whole set, separate probabilistic representations of different subsets would, in our opinion, be less likely. Yet, this approach has only provided evidence for single-point estimates (e.g., the mean) but no direct evidence for binding of feature probability distributions. Here, we aim to overcome the limitations of previous studies and test how observers encode properties of feature distributions and bind them with both spatial locations and other features.

\hypertarget{results}{%
\section{Results}\label{results}}

In three experiments, observers viewed dressed-down versions of the environment that allowed precise control of the critical aspects of feature distributions. Observers searched for an unknown oddball target that differed from other items in orientation and judged whether it was in the upper or lower half of the stimulus matrix (Figure \ref{fig:general_approach}B). Observers did this quickly and accurately despite not knowing the target or distractor parameters before each block (average response time across experiments and conditions \emph{M} = 754 ms, \emph{SD} = 197, proportion correct \emph{M} = 0.90, \emph{SD} = 0.04; see Figure \ref{fig:suppl_rawRT} for raw RT on test trials by condition).

In all experiments, the trials were organized in miniblocks of intertwined learning and test trials. In each miniblock, during five to seven learning trials, distractor stimuli were randomly drawn from two probability distributions, that were the same within each miniblock but different between miniblocks. Crucially, learning trials were organized in different ways depending on the condition. In Exp. 1, the distractors from the two distributions were either mixed together (\emph{Baseline}), colored differently (\emph{Color}) or separated into different halves of the visual field (\emph{Spatial}, see details below). On test trials, we randomly varied the similarity of the current target to non-targets from preceding trials (Figure \ref{fig:general_approach}B) with the aim of understanding how observers represent complex heterogeneous stimuli such as visual search distractors. The distractors on test trials were always from a Gaussian distribution centered at 60 to 120° relative to the current target. We assumed that during the learning trials observers encode the distractors and the distractor representation can be revealed by the response times on search trials. This would be consistent with our previous results where we have shown how response times follow the shape of the probability density function of the distractors, whether they are Gaussian or uniform, leftwards or rightwards skewed \citep{chetverikov2016Building, chetverikov2017Representing}, or even bimodal versus uniform \citep{chetverikov2020Probabilistic, chetverikov2017Rapid}. To lay the groundwork for the analyses of empirical data, we first modeled the relationship between the distractor representations and response times in a Bayesian observer model.

\begin{figure}[!t]
\centering
\includegraphics{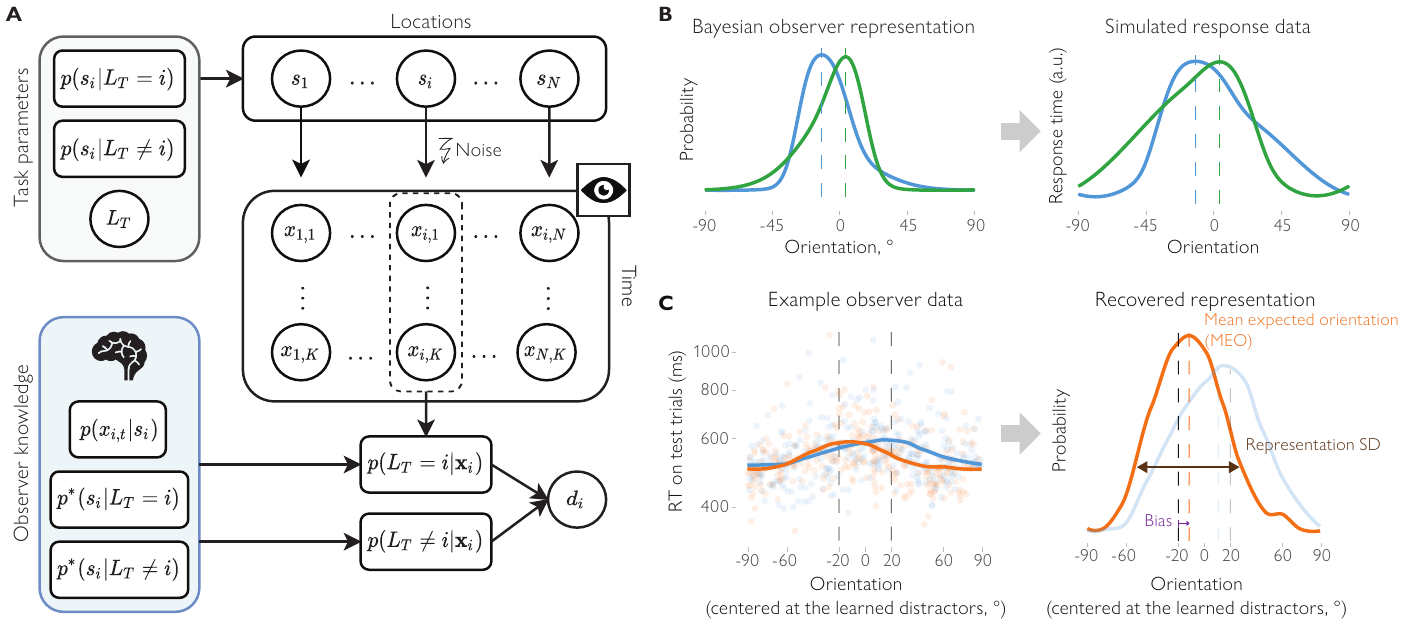}
\caption{The Bayesian observer model provides a way of reconstructing distractor representations. \textbf{A}: The Bayesian observer model. The stimuli \(s_{1}\) \ldots{} \(s_{N}\) at different locations are generated on each trial based on task parameters: the target feature distribution \(p(s_{i}|L_{T} = i\)), the distractor feature distribution, \(p(s_{i}|L_{T} \neq i\)), and the target location \(L_{T}\). At each moment in time \emph{t} within a trial and for each location \emph{i}, observers obtain samples of sensory observations \(x_{i,t}\) corrupted by sensory noise, \(p\left( x_{i,t} \mid s_{i} \right)\). Using knowledge about the sensory noise distribution and the approximation of feature distributions for targets and distractors, \(p^{*}\left( s_{i} \mid L_{T} = i \right)\) and \(p^{*}\left( s_{i} \mid L_{T} \neq i \right)\), observers compute probabilities that the sensory observations at a given location correspond to the target, \(p\left( L_{T} = i \mid \mathbf{x}_{i} \right)\), or a distractor, \(p\left( L_{T} \neq i \mid \mathbf{x}_{i} \right)\). These probabilities are combined into a decision variable \(d_{i}\) used to make a decision or to continue gathering evidence if the currently available observations do not provide enough evidence for the decision (see details in Methods). \textbf{B}: The Bayesian observer model enables predictions about response times for a given representation of distractor stimuli based on the information acquired from previous trials \(p^{*}\left( s_{i} \mid L_{T} \neq i,\bm{\uptheta}_{prev} \right)\) (see text for more details; different example distributions are shown in blue and green). Crucially, there is a monotonic relationship between the two, with response times on test trials increasing as distractor probability increases. \textbf{C}: In our analyses, we used the monotonic relationship between probabilistic representations and response times to recover the representation of distractors (right) based on the response times on test trials (left). Here, the data from an example observer in the Spatial condition is split based on whether the target was located in the left (orange) or in the right (blue) hemifield. We then estimated the parameters of the representation, such as the mean expected orientation (dashed orange line), SD and across-distribution bias (the shift in the mean towards the other distribution relative to the true mean, shown by the dashed black line).}
\label{fig:bayesian_obs_model}
\end{figure}

\subsection{Bayesian observer model} 

How do behavioral responses depend on distractor representations from previous trials? To answer this question and to reconstruct distractor representations from the behavioral responses of our observers, we built a Bayesian memory-guided observer model linking observers' internal representations of distractors to response times (Figure \ref{fig:bayesian_obs_model}A). This model is described here in short while the full description is available in the Methods.

We first describe a general structure of the model shown in Figure \ref{fig:bayesian_obs_model}A. In the model, the observer had to locate a target among a set of distractors and indicate if it was in the top or the lower part of the stimuli matrix of size $N = 36$. The features (e.g., orientation) of each stimulus \(s_{i}\) at locations \(i = 1\ldots N\) in the stimuli matrix are determined by the target location, \(L_{T}\), and the parameters of the target feature distribution, $p\left(s_{i} \mid L_{T} = i\right)$ and of the distractor feature distribution, $p\left( s_{i}|L_{T} \neq i \right)$. For each trial, these parameters are used to generate stimulus \(s_{i}\) at each location. At each moment in time \emph{t} within a trial, the observer obtains sensory samples or observations \(x_{i,t}\) at each location. These observations are not identical to the stimuli because of sensory noise that has a probability distribution \(p\left( x_{i,t} \mid s_{i} \right)\). In other words, a given stimulus might result in different sensory responses, and, conversely, a given sensory observation might correspond to different stimuli.

Crucially, we do not assume that either the task parameters, such as $p \left( s_{i}|L_{T} = i \right)$ and $p \left( s_{i}|L_{T} \neq i \right)$, or the stimuli are known to the observer. However, the observer knows the parameters of sensory noise \(p\left( x_{i,t} \mid s_{i} \right)\) and has an approximate knowledge of the target and distractor distributions denoted with asterisk, \(p^{*}\left( s_{i} \mid L_{T} = i \right)\) and \(p^{*}\left( s_{i} \mid L_{T} \neq i \right)\), that could be further separated into the knowledge based on the previous and on the current trial, e.g., \(p^{*}\left( s_{i} \mid L_{T} \neq i,\boldsymbol {\uptheta}_{prev} \right)\) and \(p^{*}\left( s_{i} \mid L_{T} \neq i,\bm{\uptheta}_{curr} \right)\) for distractors with \(\bm{\uptheta}_{prev}\) and \(\bm{\uptheta}_{curr}\) corresponding to the latent variables describing the parameters of the previous and the current trial, respectively. That is, in contrast to the traditional normative (ideal observer) models, our observer is not omniscient and does not know what was the distribution of distractors in the previous or the current trial. We assume instead that the observer has learned some approximation of the distractor distribution from previous trials and combined it with the information about the current trials to improve search efficiency.

Using this knowledge, the observer aims to find the target by comparing for each location the probability that the sensory observations are caused by a target present at that location, \(p\left( L_{T} = i \mid \mathbf{x}_{i} \right)\) against the probability that they are caused by a distractor, \(p\left( L_{T} \neq i \mid \mathbf{x}_{i} \right)\):
\begin{equation} \label{eq:1}
d_{i} = \frac{p\left( L_{T} = i \mid \mathbf{x}_{i} \right)}{p\left( L_{T} \neq i \mid \mathbf{x}_{i} \right)} 
\end{equation}
where \(\mathbf{x}_{i} = \{ x_{i,1},x_{i,2},\ldots,x_{i,t = K}\}\) are the samples obtained for location \(i\) up until a decision threshold is reached.

How can the observer estimate the probabilities in Eq. \ref{eq:1}? Here, we would focus on the distractor-related part in the denominator but similar derivations can be done for the target. First, following the Bayes rule, the posterior probability that a distractor is at a given location is proportional to the likelihood of samples being drawn from the distractor distribution:
\begin{equation} \label{eq:2}
p\left( L_{T} \neq i \mid \mathbf{x}_{i} \right) \propto p\left( \mathbf{x}_{i} \mid L_{T} \neq i \right)
\end{equation}

Assuming that the samples are independent in time their probability in log-space is equal to a sum of individual probabilities:
\begin{equation} \label{eq:3}
\log{p\left( \mathbf{x}_{i} \mid L_{T} \neq i \right)} = \sum_{t=1}^{K} \log{ p\left( x_{i,t} \mid L_{T} \neq i \right) }
\end{equation}

Importantly, the probability of a single observation corresponding to a distractor can be found by integrating over all possible stimuli values:
\begin{equation} \label{eq:4}
p\left( x_{i,t} \mid L_{T} \neq i \right) = \int_{}^{}{p\left( x_{i,t} \mid s_{i} \right)p^{*}\left( s_{i} \mid L_{T} \neq i \right)}ds_{i} 
\end{equation}
In other words, the observer combines the knowledge about sensory noise and the distractor distribution to estimate that a given sensory observation corresponds to distractors.

Notably, we are mainly interested in the test trials where the parameters of the current trial are independent of the parameters of the previous trials, hence:
\begin{equation} \label{eq:5}
p^{*}\left( s_{i} \mid L_{T} \neq i \right) \propto p^{*}\left( s_{i} \mid L_{T} \neq i,\boldsymbol{\uptheta}_{prev} \right) 
\end{equation}

To reiterate, in our experiments, by design, the parameters of the current trial are controlled with respect to the current stimuli (i.e., the distractors on the current test trial are drawn from a distribution with a mean from 60° to 120° off the current test target). Hence, only \(p^{*}\left( s_{i} \mid L_{T} \neq i,\boldsymbol{\uptheta}_{prev} \right)\) matters for relative changes in response times.

Notably, if sensory observations are obtained with high frequency and sensory noise is low relative to the uncertainty in distractor representations, the log-sum of probabilities can be approximated as:
\begin{equation} \label{eq:6}
\sum_{t = 1}^{K}{\log{ p\left( x_{i,t} \mid L_{T} \neq i \right) }} \approx K\left( \log{p^{*}\left( s_{i} \mid L_{T} \neq i  \right) + C }\right) 
\end{equation}
In words, if many samples are acquired at a given location, the log-probability that they are caused by a target is approximately equal to the number of samples obtained, times the log-probability that a stimulus at this location is a distractor.

Using Eq. \ref{eq:1}, \ref{eq:5} and \ref{eq:6} (and analogous equations for target representations), the observer then can estimate the decision variable for all the samples obtained as:
\begin{equation} \label{eq:7}
\log d_{i} \propto K\left( \log{p^{*}\left( s_{i} \mid L_{T} \neq i,\boldsymbol{\uptheta}_{prev} \right)} - \log{p^{*}\left( s_{i} \mid L_{T} \neq i,\boldsymbol{\uptheta}_{prev} \right)} \right)
\end{equation}
with constants subsumed under the proportionality sign. In words, the decision variable when a decision is made is proportional to a difference in the amount of evidence that a stimulus is a distractor and that it is a target times the number of samples obtained.

Finally, assuming that target and distractor representations are independent and noting that the response time is proportional to a number of observations needed to reach a decision, \(RT \propto K\), the observer representation of distractor features learned from previous trials is related to response times:
\begin{equation} \label{eq:8} 
RT \approx \frac{C_{1}}{C_{0} - \log{p^{*}\left( s_{i} \mid L_{T} \neq i,\boldsymbol{\uptheta}_{prev} \right)}}\ 
\end{equation}
where \(C_{0}\) and \(C_{1}\) are constants (see details in Methods). In words, there is an inverse relationship between response times and the approximate likelihood that a given stimulus is a distractor, \(p^{*}\left( s_{i} \mid L_{T} \neq i,\boldsymbol{\uptheta}_{prev} \right)\), with the information obtained from previous trials described by a set of latent parameters, \(\boldsymbol{\uptheta}_{prev}\).

While this decision model is relatively simple, it provides a good intuition for observer behavior in the task (a more optimal model is provided in the Supplement but the conclusions do not depend on model choice). The model does not make any assumptions of how the observer learns these parameters. However, it shows that when the probability that a stimulus at a given location (e.g., a test target) is a distractor is lower, response times are lower as well, and vice versa.

This model provides an important insight, namely, that observers' representations are monotonically related to response times (Figure \ref{fig:bayesian_obs_model}B). Hence, the monotonic relationship between the distribution parameters (mean, standard deviation, and skewness) reconstructed from RTs and from the true representation parameters would hold under any other monotonic transformation (for example, if RTs are log-transformed and the baseline is subtracted as we do in our analyses; see also Figure \ref{fig:sim_simpl_IO}). In other words, the analysis above shows how response times can be used to approximately reconstruct observers' representations of distractors and estimate their parameters.

\subsection{Binding orientation probabilities to locations and colors}

Having shown how observer response times can be related to the distractor representations, we now turn to the empirical data. Observers' response times to different test targets allow us to infer which orientations were the most difficult to find, resulting in the longest response times. As explained above, this methodology has enabled us to reveal how observers can represent distractor distributions in surprsising detail \citep{chetverikov2016Building,chetverikov2017Representing,chetverikov2020Probabilistic}. Crucially, we can then reconstruct observers' representations of the probability distributions during learning trials (see Methods).

The experiments differed in the structure of the learning trials. There were three conditions in Experiment 1. The learning trials in the \emph{Spatial} condition were organized so that distractor distributions in the left and the right hemifield differed to mimic the clustering of similar visual stimuli in the real world. In the \emph{Color} condition, instead of spatial grouping, different distractor subsets were grouped by color while individual items were randomly distributed. Finally, in the \emph{Baseline} condition, items from the two distributions had the same color and were randomly distributed (Figure \ref{fig:general_approach}C).

Firstly, we report the results on the mean expected orientations (MEO) corresponding to the means of the recovered representations (Figure \ref{fig:bayesian_obs_model}C). If observers ignore the separation of the two parts of the distribution, then MEO should match the mean of the overall distribution, but should differ between the distributions if the representations are bound to locations or colors. For example, if observers accurately learn the properties of the distributions, the MEO should be at +20° relative to the overall mean in the Spatial condition when the test line is presented in the hemifield that previously had distractors with an average relative orientation of +20°.

\begin{figure}[!bt]
\centering
\includegraphics[width=\textwidth,height=\textheight,keepaspectratio]{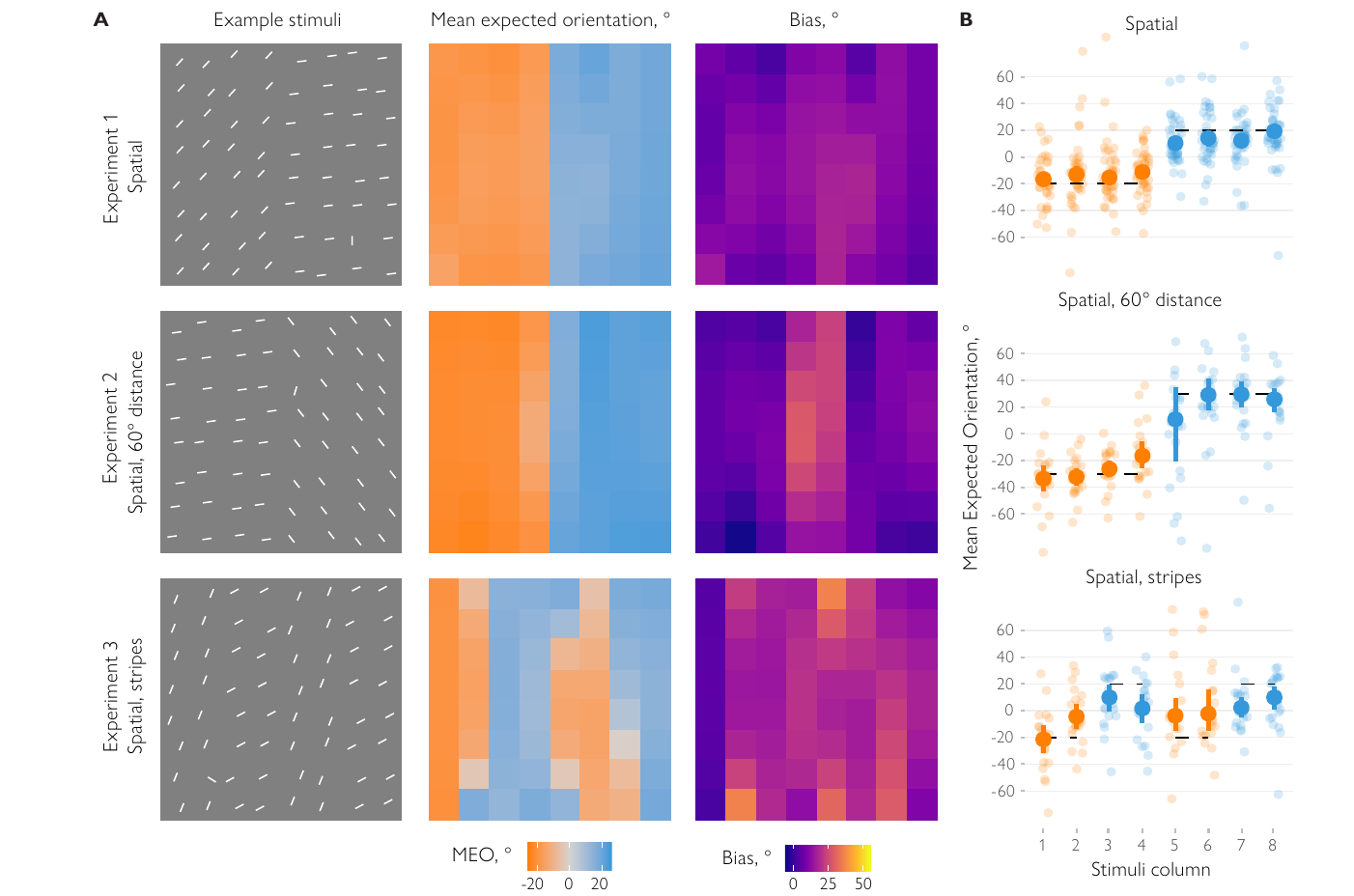}
\caption{Spatial structure of probabilistic representations. \textbf{A}: Example stimuli (left column), recovered mean expected orientations (middle column) and the across-distribution biases in mean expected orientations relative to the true orientations at a given location (right column). The stimuli show a single learning trial from the search task in the corresponding experiment. The mean expected orientation (MEO) was then computed at each location relative to the overall average orientation in the preceding learning block. For presentation purposes, the data were rearranged so that the distribution in the left hemifield (or in the columns 1,2,5,6 in the stripes condition) was oriented clockwise relative to the overall mean. The biases in MEO were computed by subtracting the mean orientation for a given part of the distribution (e.g., at the left/right hemifield in the Spatial condition of Experiment 1) and recoding the resulting errors so that the positive values correspond to a bias towards the other distribution. \textbf{B}: Average MEO by column of stimuli matrix in the spatial conditions. Small dots show the data for individual observers, larger dots and bars show means and 95\% CI, respectively. Dashed horizontal lines show the true means for a given part of the distribution.}
\label{fig:meo_maps}
\end{figure}

We found that in the Spatial condition, observers' representations in each hemifield followed the actual physical distractor distribution (Figure \ref{fig:meo_maps}). The estimated MEO relative to the overall mean was \emph{M} = -14.02° (\emph{SD} = 6.02) and \emph{M} = 14.90° (\emph{SD} = 5.14) for probes for clockwise (CW) and counterclockwise-shifted (CCW) distributions, respectively. The difference in MEO between the two distributions was much greater than zero (\emph{b} = 28.94°, 95\% HPDI = {[}25.34, 32.56{]}, \emph{BF} = \SI{6.35e17}) showing that observers expected different orientations in different hemifields. We then computed the across-distribution bias by recoding the errors in MEO relative to the true mean for each distribution so that positive values correspond to shifts towards the other distribution. That is, the bias here represents by how much observers' expectations deviated from the true mean orientation at a given location towards the mean orientation at the other location. For both hemifields there was a significant bias towards the other hemifield (\emph{M} = 5.52°, 95\% CI = {[}1.86, 9.14{]}). This shows that while observers represent the spatial separation between the two distributions, signals from the other hemifield still influence their responses.

But does spatial separation help observers track the feature probabilities? In the Baseline conditions, locations of the CW and CCW distributions were chosen randomly for each learning trial. We repeated the analysis described above, comparing the response to test targets at the location that had CW and CCW orientations on the immediately preceding trials. We expected to find stronger across-distribution biases as there was no separation between the distributions across trials. Importantly, the across-distribution bias was larger in the Baseline (bias \emph{M} = 11.35°, 95\% CI = {[}7.71, 15.00{]}; Figures \ref{fig:meo_maps} and \ref{fig:higher_order_pars}) than the Spatial condition (effect of condition \emph{M} = 5.84, 95\% CI = {[}1.10, 10.58{]}, \emph{BF} = 108.24). In other words, the representations for each distribution were closer to the overall distribution mean in the Baseline than the Spatial condition. This argues that when the learned distributions are consistently presented at separate locations, observers can track them better than when they are randomly distributed.

Do observers integrate information about orientation probabilities and color? In the Color condition, the locations of the test targets were counterbalanced with respect to their colors, so we should only find differences in MEO if observers formed an association between color and orientation. Indeed, we found that the MEOs for the two distributions differed (\emph{b} = 7.35, 95\% HPDI = {[}1.30, 13.06{]}, \emph{BF} = 148.04) although across-distribution biases were stronger (\emph{M} = 16.30, 95\% CI = {[}12.66, 19.86{]}) than in the Spatial condition (the difference between conditions \emph{M} = 10.78, 95\% CI = {[}5.99, 15.54{]}, \emph{BF} = \SI{6.56e4}). This means that if observers saw yellow lines shifted CW and blue lines shifted CCW relative to the overall distractor mean during learning trials, they learned this association which affected their response times on subsequent test trials. Importantly, this demonstrates that observers can integrate information about likely orientations with information about other features (in this case color), even if there is no spatial information to guide this integration.

\subsection{Encoding orientation probabilities at different spatial scales} 

Having established that observers associate information about most likely orientations with specific locations or colors, we then asked if we can uncover the origins of the observed biases by assessing the recovered representations in the Spatial condition in more detail (for this and later analyses, we increased the sensitivity of our analyses by combining the data from the Spatial group in Experiment 1 with an additional participant group that performed the same task; see Methods). We computed MEO using the aggregated data from all participants for each location in the stimuli matrix in this condition. As Figure \ref{fig:meo_maps} shows, across-distribution biases were stronger closer to the boundary between the two hemifields. We then tested this observation by directly comparing MEOs for test trials with targets presented at the boundary (two central columns) between the hemifields against other test trials. We found that the bias was significantly larger at the boundary between the two distributions than in the other columns (\emph{M} = 4.80° (\emph{SD} = 6.99) and \emph{M} = 9.04° (\emph{SD} = 11.36), \emph{b} = 4.23, 95\% HPDI = {[}0.21, 8.32{]}, \emph{BF} = 42.34; Figure \ref{fig:meo_maps}B). However, the biases were also significantly above zero outside the boundary (\emph{BF} = 248). This suggests that the distribution representations are not homogeneous and influence each other strongly when they are close in space, but this mutual influence also extends outside the immediate neighboring locations (see Discussion).

\subsection{Bias strength depends on similarity and spatial arrangement} 

In two follow-up studies, we further investigated observers' representations of spatially-grouped heterogeneous stimuli. In Experiment 2, we tested whether the similarity between the distributions along the tested feature dimension (orientation) affects the strength of the across-distribution biases. Recent studies suggest that similarity is an important factor determining whether the information is pooled or not at different levels of perceptual processing \citep[e.g.,][]{coen-Cagli2015Flexible, herrera-Esposito2021Flexible, manassi2012Grouping, qiu2013Segmentation, utochkin2018Continuous}. We hypothesized that the bias should be stronger when the stimuli from the two distributions are more likely to have the same cause in the external world. For example, the boundary effect in Experiment 1 might occur because stimuli that are close in space are more likely to belong to the same object. By the same reasoning, if the two distributions are less similar, they are less likely to have the same cause, and the biases should be weaker.

To test this, we used the same spatial arrangement as in the Spatial condition in Experiment 1, but the distribution means were now 60° away from each other instead of 40° as in Experiment 1 (see example stimuli in Figure \ref{fig:meo_maps}A). We found that again, MEO's were close to their true values with \emph{M} = 26.35° (\emph{SD} = 13.43) and \emph{M} = -27.65° (\emph{SD} = 10.65) for distributions centered at 30° and -30° relative to the overall mean, respectively. Importantly, while there was a strong bias at the boundary between the distributions, \emph{M} = 19.05° (\emph{SD} = 27.27), \emph{BF} = 8.36, it was absent at other positions (bias \emph{M} = 0.60° (\emph{SD} = 8.65), with \emph{BF} = 4.12 in favor of no bias). Experiment 2, therefore, shows that reducing the similarity between the distributions eliminates the biases except for the immediately adjacent locations.

In Experiment 3, we tested whether an even more complex spatial arrangement would allow us to recover the ``map'' of observers' expected orientations. To this end, the stimuli were organized in ``stripes'' of two matrix columns with two different distributions from Experiment 1 (with means separated by 40°) positioned at odd and even stripes (counterbalanced across blocks, Figure \ref{fig:meo_maps}A). We found that observers expected clockwise-rotated orientations (\emph{M} = 6.20°, \emph{SD} = 9.91) at locations of stripes rotated 20° clockwise relative to the overall mean and counterclockwise-rotated orientations (\emph{M} = -11.04°, \emph{SD} = 17.11) at other stripe locations. However, the across-distribution bias (\emph{M} = 11.70°, \emph{SD} = 7.52) was stronger than in the Spatial condition in Experiment 1 (\emph{b} = 5.90, 95\% HPDI = {[}2.50, 9.33{]}, \emph{BF} = 4.30). This demonstrates that while separating distributions in space helps observers track distributions (as shown in Experiments 1 and 2), the effects of spatial organization decrease as the organization becomes more complex.

\begin{figure}[bt]
\centering
\includegraphics[width=\textwidth,height=\textheight,keepaspectratio]{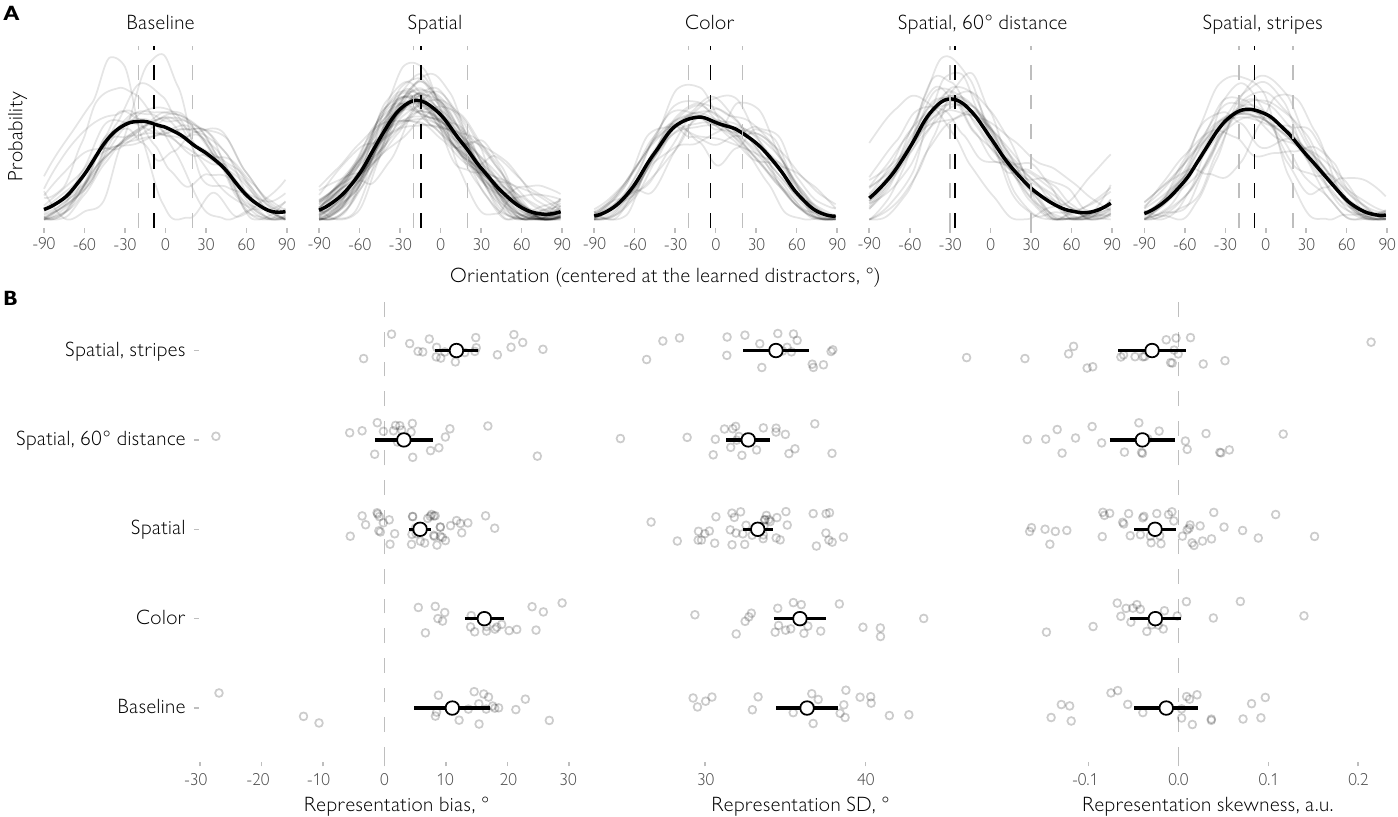}
\caption{Recovered average representations and their parameters across experiments and conditions. \textbf{A}: The black curves show the average representation while representations for individual observers are shown in light gray. Dashed vertical lines show the mean of the representation (black) and the true mean of the stimulus distributions (light gray). Note that the representations are aligned so that when two distributions are present, the true mean at the tested location is clockwise (-20° or -30°) while the other mean is counterclockwise (20° or 30° relative to the true mean). \textbf{B}: Estimated parameters (bias, SD and skewness). Large dots and errorbars show the mean across observers for a given parameter and the associated 95\% confidence intervals. Smaller dots show data for individual subjects.}
\label{fig:higher_order_pars}
\end{figure}

\subsection{Higher-order parameters of probabilistic representations} 

Next, we asked whether observers' representations contain more information about the distributions than just their average? We used the reconstructed distractor representations (Figure \ref{fig:higher_order_pars}A) to estimate their circular standard deviation and circular skewness. The former corresponds to the expected variability among distractors, while the latter quantifies their symmetry.

First, we hypothesized that if the variability of the distributions is encoded, then the expected variability would be higher when the distractor distributions are less well separated. Indeed, we found that observers' expectations about distractor variability differ between conditions (\emph{BF} = \SI{2.03e5}) with lower SD when the distractors were separated by hemifields (\emph{M} = 33.3, 95\% HPDI = {[}32.2, 34.4{]} for the Spatial condition with 40° separation and \emph{M} = 32.7, 95\% HPDI = {[}31.1, 34.2{]} for 60° separation) compared to other conditions (\emph{M} = 35.9, 95\% HPDI = {[}34.4, 37.5{]} in the color condition, \emph{M} = 34.4, 95\% HPDI = {[}32.9, 35.9{]} for the stripes arrangement condition). When the two distributions were less well separated, observers were more uncertain in their estimates, leading to distractor representations with higher SD's (Figure \ref{fig:higher_order_pars}B).

We also expected that the distribution presented at the tested location or in the tested color would weigh more highly in the resulting representation, causing an asymmetry. Alternatively, if observers only use the mean and variance to encode the distribution (as assumed by ``summary statistics'' accounts), then the represented distribution should be symmetric. We found that observers' representations were asymmetric in all conditions, with a higher probability mass at the side corresponding to the distribution presented at the tested location or in the tested color, \emph{M} = -0.03, 95\% CI = {[}-0.04, -0.02{]}. Notably, however, no differences between conditions were found, \emph{BF} = \SI{1.99e-6}, indicating that symmetry is not affected by the way the distributions are organized in the display. In sum, observers represent not only the average stimulus values but also their variability, and the representations are skewed towards distributions presented at other locations or in different colors.

\hypertarget{discussion}{%
\section{Discussion}\label{discussion}}

Our main hypothesis was that observers extract information about probabilities of visual features from heterogeneous stimuli and bind the resulting probabilistic representations with locations on the one hand and other features on the other. Our results support both these proposals very clearly. Importantly, this demonstrates for the first time, how the visual system can build probabilistic representations of the visual world by extracting information about the features of complex heterogeneous stimuli.

A visual search task allowed us to uncover representations of heterogeneous distractors. We formulated a Bayesian observer model and demonstrated analytically and through simulations that response times are a monotonic function of observers' expectations about distractor orientations, supporting earlier empirical findings \citep{chetverikov2016Building, chetverikov2019Feature, chetverikov2020Probabilistic, chetverikov2017Set, hansmann-Roth2021Dissociating, hansmann-Roth2019Representing, tanrikulu2020Encoding, tanrikulu2021Testing}. Using this knowledge, we were able to estimate the characteristics of observer representations -- their means, precision, and skewness -- and to assess how they vary depending on whether observers can associate them with locations or with other, task-irrelevant features, such as color.

We found that observers both encode and combine feature distributions in scenes containing two different distributions. The representations generally follow the physical distribution of the stimuli for a given location or a given color, but importantly, observers are also biased towards the other distribution. The strength of the bias depends on the degree of separation between the distributions. When the distributions were separated in space, observers' representations of one distribution were less influenced by the other distribution, compared to when they were separated by color or were intermixed (Baseline condition). Furthermore, as we found in Experiment 3, more complex spatial arrangements (``stripes'') increased the biases towards the other distribution. In sum, observers bind probabilistic representations of visual features to locations and other features, but such binding is not impenetrable, reminiscent of `illusory conjunctions' of discrete feature values \citep{treisman1982Illusory}.

We were then able to recover the representation of the distribution at different spatial scales. We found that for spatial separation, the biases are stronger at the boundary between the two distributions. This is reminiscent of the hierarchical organization of information about feature probabilities within a scene proposed for perceptual ensembles \citep{alvarez2011Representing, haberman2012Ensemble}. Such hierarchical ensemble models suggest that observers represent information about feature probabilities at different levels: for example, the orientation statistics at a particular location are combined to form a representation for a group of items, which are, in turn, combined to form an overall ensemble representation. Our results agree with this idea: the stimuli observers expect at a given location depend not only on what was previously shown at this location but also on stimuli presented at other locations. Crucially, biases were also present for the Color condition as well as for the non-boundary locations in the Spatial condition of Experiment 1. This indicates that the results cannot be explained by purely local summation of the inputs. It remains to be tested whether there are actual separable representations of probability distributions at different levels, or just a unified spatio-featural map guiding observer responses.

We hypothesized that the representations should be more biased by each other when they are more likely to have the same cause in the external world. This could provide a normative explanation for the boundary effect: sensory input from adjacent locations is likely to be caused by the same object and should therefore be integrated while locations far away from each other should be treated separately. Similarly, for example, in multisensory integration studies, auditory and visual signals are less likely to be integrated when there is a large discrepancy in their locations \citep{kording2007Causal, shams2010Causal}. However, in Experiment 1, we found across-distribution biases at locations far from the other distribution. We reasoned that this is because the stimuli themselves are similar enough to be potentially caused by the same object, and the inputs are therefore integrated even from non-neighboring locations. In Experiment 2, we tested this explanation by asking if the similarity between the distributions themselves in the tested feature domain (orientation) also plays a role. We found that when the distributions were made more dissimilar, the biases were observed only at the boundary between the distributions but not at other locations. That is, observers no longer take into account the input from non-neighboring locations, when stimuli are dissimilar. Speculatively, introducing longer learning streaks could also help to reduce the bias by increasing precision of the representations \citep{chetverikov2017Rapid}. This supports the proposed normative explanation and suggests that the principles of information integration for heterogeneous visual inputs are the same as for other cases, such as multisensory integration or estimation of complex visual features \citep{landy2011Ideal}.

We then tested if observers represent more than just the mean distractor orientation. We found that observers represent the distractor variability (i.e., the standard deviation or width of their representations), which varies in a predictable fashion with the separability between distractor distributions. When distractor distributions are poorly separated (e.g., only by color or are organized in `stripes'), their representations are wider, indicating more uncertainty. Furthermore, the representations are asymmetric where the tail of the distribution corresponding to the orientations matching the tested location or color is fatter. While we are agnostic to the specific mechanisms of how the information is integrated across different parts of the visual field, speculatively, such asymmetric distributions can be seen as an output of a hypothetical weighting process. As a computational abstraction, the resulting representation can be seen as a normalized sum of basis functions (similar to a kernel density estimator). The weight of a certain basis function could depend on how well it matches the stimuli across the visual field (i.e., how many distractors had a certain orientation) and their relevance to the current goals. The presence of skewness indicates that observers do not simply represent the distractors with a (biased) mean and variance, their representations have a complex shape with more relevant information (e.g., previous orientations at a tested location) weighted higher and less relevant information (e.g., previous orientations at the other locations) having lower weight, but still influencing the outcome. However, we do not see the effect of condition on the distribution asymmetry, which would be expected if the matching and nonmatching parts of the distribution were combined as a weighted mixture with weights depending on the degree of separation. The absence of the condition effect might be related to the greater difficulty of precisely estimating the amount of skewness as opposed to mean or variance. Nevertheless, the overall skew in the representations is indicative of how sophisticated the learning can be where various factors, such as the amount of information about the underlying probability distribution and task-relevance of the stimuli in each case, are taken into account in the representation, and how this determines how different parts of the display are weighted.

These findings indicate that observers represent information about distractor features as a probability distribution rather than only in terms of the summary statistics, in contrast to popular ideas of simple ``summary statistics''. For example, Treisman \citeyearpar{treisman2006How} argued that statistical processing is a distinct mode of perceptual and attentional analysis of stimulus sets. She proposed that because of limited attentional capacity statistical summaries are generated that include the mean, variance, and perhaps the range. These summaries enable rapid assessment of the general properties and layout of natural scenes \citep{chong2005Statistical, emmanouil2008Dividing}. Similarly, Rahnev \citep{rahnev2017case, yeon2020suboptimality} argued that observers represent only a summary consisting of the most likely stimulus and the associated strength of evidence, and Cohen et al. \citeyearpar{cohen2016What} used summary statistics to explain the richness of consciousness experience. Our results argue against such views, since the representations that are bound together are far more detailed than this implies. That is, the brain might instead approximate the visual input by using a complex set of parameters to provide accurate descriptions of feature probabilities \citep{freeman2011Metamers, rosenholtz2020Demystifying}.

A recent finding may provide insights into why summary statistic accounts have been so poular. Hansmann-Roth et al. \citeyearpar{hansmann-Roth2021Dissociating} reasoned that optimal behavior requires the encoding of full feature distributions, not only summaries, but observers might be unable to explicitly report the full distribution. This is analogous to how difficult it might be to verbally describe the variety of colors of an apple without resorting to simplifications (see Figure \ref{fig:general_approach}A). Hansmann-Roth et al. tested observers' representations both implicitly and explicitly and while explicit judgments were limited to the mean and variance of feature distributions, implicit measures revealed detailed representations of the same distributions. More information was therefore available to observers than studies of summary statistics, that have mostly relied on explicit measures, have indicated. Crucially, Hansmann-Roth et al. were able to uncover why this is: revealing these detailed representations requires implicit methods where representations are probed by assessing the effects of role reversals between targets and distractors as we do here.

Can the results observed in our study be explained by simple sensory adaptation mechanisms? Sensory adaptation is a well-studied phenomenon: at a neural level, exposure to a stimulus alters neural responses to subsequently presented stimuli, while behaviorally adaptation often results in a repulsive bias with estimates of, for example, orientation in the adjustment task shifted away from the adapter \citep{clifford2007Visual, schwartz2007Space}. In our task, observers were exposed to a certain distractor distribution within each miniblock, so their behavior could be influenced by adaptation. We tested this idea by developing a variation of a model reported here assuming that the observer does not encode the knowledge about the distractors directly (\(p^{*}\left( s_{i} \mid L_{T} \neq i,\bm{\uptheta}_{prev} \right)\) is flat) but instead the sensory space is warped to efficiently encode the distractor distribution \citep{stocker2005Sensory, wei2015Bayesian, wei2012Efficient}. The results (not shown) indicated that the search would be the most efficient when a target matches previous distractors, contrary to our findings and the well-known role-reversal effects in visual search literature \citep{kristjansson2008Priming}. The intuition here is that when targets and distractors are on the opposite sides of the adapter, they are actually pushed together due to the circularity of orientation space. Furthermore, recent studies using a similar behavioral task by \cite{rafiei2021You,rafiei2021Optimizing} and \cite{pascucci2022Feature} also speak against the involvement of adaptation. Rafiei et al. looked at how the perception of a current search target \citeyearpar{rafiei2021Optimizing} and a neutral line \citeyearpar{rafiei2021You} is affected by distractors and previous targets. Importantly for the current issue, they found that when distractors are similar to the test item, its estimates are pulled towards the distractors, while the adaptation profile is generally repulsive for similar items. In a recent study by Pascucci, Ceylan, and Kristjánsson \citeyearpar{pascucci2022Feature}, the authors found that when observers passively viewed an array of lines that all came from the same distribution (no singleton), no learning of the distribution occurred. In summary, both the modelling results and the empirical data suggest that sensory adaptation cannot explain our findings.

Where does the uncertainty in the distractor representation come from and how the observers learn about it? Our study is agnostic on this issue. We do not assume any parametric form for this representation (that is, we do not assume that observers represent, for example, the mean or variance). Note, however, that it is unlikely that a single-trial representation is a simple Gaussian and only when aggregating across trials the skewness appears. This idea was tested in a recent study by Chetverikov et al. \citeyearpar{chetverikov2020Probabilistic}, where observers have to find two targets on each trial with a mixture distribution similar to the Baseline condition used here. By comparing a model assuming that the distribution is approximated as a single-peaked and a model assuming that the distribution is approximated as a mixture of two parts, we found that the hypothesis of a simple Gaussian representation is not supported by the data. It might well be, however, that at a more detailed level, for example, at a level of a single location or a single moment in time, the representation is a simple one. Yet, from the computational perspective, these simpler representations have to be combined into an aggregated representation of distractors shown here and in our previous studies to be corresponding to a probability distribution of distractors with more details that the simple summary statistics would allow.

In our experiments, observers learn the distractor feature by combining inputs from heterogeneous stimuli across several trials in each block, and it can be argued that this is different from perceiving a single stimulus on a single trial. However, the visual cortex aggregates information on many different timescales \citep{delange2018How}. Even on a single trial, perception unfolds in time and at each moment is dependent on what has been seen before. And even for a simple stimulus, the visual cortex receives inputs from many retinal neurons that are affected by processing noise, potentially indistinguishable from the input from varying features. Indeed, this is why stimulus variability (`external noise') is often used to manipulate visual uncertainty \citep{barthelme2009Evaluation, henaff2020Representation}. We therefore believe that distinguishing ``simple'' and ``complex'' perception is impossible. However, our results clearly show that information about feature probabilities is available for visually-guided behavior.
\section{Summary}
Taken together, our results show that observers can not only encode probabilities of features from heterogeneous stimuli in detail, but also integrate them with both locations and other features that have different distributions. These results arguably represent the strongest support yet for the long-standing idea that the brain builds probabilistic models of the world \citep{chetverikov2017Learning, fiser2010Statistically, knill2004Bayesian, orhan2015Neural, rao2002Probabilistic, sahani2003Doubly, tanrikulu2021What} and show that probabilistic representations can serve as building blocks for object and scene processing. Notably, such representations are not simply limited to summary statistics (e.g., a combination of mean and variance; \cite{cohen2016What}). Our results also indicate that observers do not represent physical stimuli precisely, but instead construct an approximation influenced by input from other stimuli. This probabilistic perspective stands in sharp contrast to views where discrete features of individual stimuli are \emph{either} bound together to form objects or processed ``statistically'' \citep{rosenholtz2020Demystifying, treisman2006How}. Instead, we suggest that the probabilistic representations are automatically bound to locations and other features since such binding occurred even though it was not required in the task. Probabilistic representations are therefore not acquired in isolation but constitute an integral part of perception.

\hypertarget{methods}{%
\section{Methods}\label{methods}}

\subsection{Participants}
In total, eighty observers (fifty female, age \emph{M} = 23.10) participated in the experiments. Twenty observers (ten female, age \emph{M} = 25.45) participated in the first experiment (Baseline, Spatial, and Color conditions) split across two sessions. Twenty observers (fourteen female, age \emph{M} = 25.00) participated in Experiment 2 (``Spatial, 60° distance'') and another twenty (thirteen female, age \emph{M} = 25.45) in Experiment 3 (``Spatial, stripes''). Finally, the data from additional twenty observers (thirteen female, age \emph{M} = 16.50) were collected for the Spatial condition of Experiment 1 to increase the sensitivity of the spatial analyses.

All were staff or students at the Faculty of Psychology, St.~Petersburg State University, Russia, or the University of Iceland, Iceland. The experiment was approved by local ethics boards and was run in accordance with the Helsinki declaration. Participants at St.~Petersburg State University were rewarded with 500 rubles (approx. 8 USD) per hour each, participants at the University of Iceland participated without additional reward. All gave their informed consent before participating. The participants were naïve to the purposes of the studies. Participants were given ample time for training until they felt comfortable doing the task (the training time ranged from 5 minutes to one hour depending on the participant).

\subsection{Procedure} 
In \emph{Experiment 1,} each participant performed a search task in five conditions. In each condition on each trial, observers were presented with 8×8 matrices of 64 lines (line length: 0.71° of visual angle; matrix size: 16×16°; uniform noise of ±0.5° was added to each line coordinate). The goal was to find the odd-one-out line whose orientation differed most from the others. Sessions were divided into miniblocks of 5 to 7 learning trials followed by 1 or 2 test trials (the number of trials chosen randomly for each block; the variation in the number of trials was introduced to decrease the effect of temporal expectations, \cite{shurygina2019Expectations}. During learning trials, the overall mean of distracting items stayed the same within each miniblock (but varied randomly between miniblocks) with half of the distractors drawn from one distribution and the other half from another distribution with the properties of distributions differing between conditions:

\emph{Baseline}: two truncated Gaussian distributions with SD = 10° and range of 40°, with means separated by 40° (±20° relative to the overall mean), all stimuli had the same color (white); half of the distractors were drawn randomly from one distribution, half from another, then they were positioned randomly within the stimuli matrix and then a randomly chosen distractor was replaced with a target.

\emph{Spatial}: two distributions (either a truncated Gaussian with SD = 10° and a range of 40° or uniform with the range of 40° in random combinations) with means separated by 40° (±20° relative to the overall mean), all stimuli had the same color (white), one distribution was shown in the left half of the matrix, the other in the right half.

\emph{Color}: the same distributions as in the Spatial condition were used, but lines drawn from one distribution were blue, while lines from the other distribution were yellow. Positions for each line within the stimuli matrix were chosen randomly.

In all cases, two lines were added to each distractor distribution with their orientation equal to the minimal and maximal values from that distribution range. As a result, Gaussian and uniform distributions always had the same range. The target orientation on each trial was drawn randomly from a uniform distribution ranging between 60° and 120° relative to the mean distractor orientation.

On test trials, distractors came from a single Gaussian distribution with SD = 10° (range-restricted in the same way as described above) and a mean sampled from a 0-180° uniform distribution, while target orientation was determined in the same way as on the prime trials (that is, selected randomly from 60 to 120° range relative to the current distractor mean). In the color condition, half of the lines from that distribution were blue, half were yellow.

The Baseline condition had 2304 trials, while the Spatial and Color conditions had 5376 trials each with the higher number of trials used in the latter case to counterbalance additional factors (distribution type combinations). The trials were split into two (for Baseline) or four sessions (other conditions) with a break for rest halfway within each session. Observers participated in each session at a separate time depending on availability but with a break of no less than two hours between sessions and no more than two sessions within a day. The order of sessions with respect to conditions was counterbalanced with a Latin square design.

\emph{Experiments 2 and 3} followed the same general procedure as the Spatial condition of Experiment 1. In Experiment 2 the means of the distributions were separated by 60° (±30° relative to the overall mean) instead of 40° in Experiment 1. In Experiment 3, the two distributions were separated by 40°, as in Experiment 1, but arranged in ``stripes'' so that the lines drawn from the first distribution were positioned in the 1\textsuperscript{st}, 2\textsuperscript{nd}, 5\textsuperscript{th}, and 6\textsuperscript{th} columns of the stimuli matrix while the other columns were populated with lines from the second distribution. In both experiments, each participant took part in two sessions of 1536 trials each with a rest period halfway within each session.

\subsection{Data processing} 
For our main analyses of interest, incorrect responses were excluded, and response times were log-transformed and centered by subtracting the mean for each participant. Then, to reduce the noise in RT measurements, spatial and featural confounders were removed (the results remain similar when no corrections are applied). First, the effect of the distance between target locations on consecutive trials and the effect of the target location were removed by regressing out the fifth-degree polynomials of the absolute distance (in degrees of visual angle) between the target locations on the current and the previous trials and the current targets horizontal and vertical coordinates. Then, we also removed potential influences from the well-known oblique effect (the search speed differs between oblique and cardinal stimuli \cite{chetverikov2017Learning, wolfe1999Which} by regressing out the fifth-degree polynomials of target and distractor obliqueness computed as an absolute distance in degrees to the nearest cardinal orientation. The regression was run separately for each experiment and condition.

To reconstruct observers' distractor representations, we used the response times on the first test trial in each miniblock. We then converted the response times as a function of the similarity between the test target and the previous distractor mean to a probabilistic representation and estimated its parameters.

To convert the noisy response times into probabilities, we first smoothed RT as a function of the test target and previous distractor mean using the local regression approach (a generalization of the moving average) for each observer in each condition. To account for circularity, we appended 1/6 of the data from each end of the orientation space to the opposite end before smoothing. In analyses applied to each stimulus location, we further assumed that RTs are a smooth function of the stimuli matrix row within the local regression while columns of the stimuli matrix were treated independently. We then transformed a smoothed RT function into a probability mass function by subtracting the baseline and normalizing to one. Finally, we computed the parameters of the recovered probabilistic representation: the mean expected orientation (circular mean), circular standard deviation, and circular skewness as defined by Pewsey \citep{pewsey2004Large}. Note that under the hypothesized Bayesian observer model, the estimated standard deviation and skewness are monotonically related to the true parameters of the distractor representation but are not identical to it (additionally confirmed in simulations, Figure \ref{fig:sim_simpl_IO}).

Unless stated otherwise, we used Bayesian hierarchical regression with the \emph{brms} \citep{burkner2017brms} package in R. Note that while we include Bayes factor values in the description of the results, we were mostly interested in measuring the effects of the variables of interest in our models, and hence the models included the default flat (uniform) priors for regression coefficients. Given that Bayes factors are prior-dependent, we believe that the information provided by the 95\% highest-density posterior intervals (HDPI) is more useful for judging the results than the Bayes factors. To make sure that the conclusions are not dependent on the particular analytic approach, we repeated the analyses using the conventional frequentist statistical test with the same results (the report using this approach is provided alongside the data in an online repository, see \hyperlink{data-availability}{data availability} statement).

\subsection{Bayesian observer model}
In our experiments, participants located a target among a set of distractors and indicated if it was in the upper or the lower part of the stimuli matrix. On each trial, the experimenter sets the task parameters, namely, parameters of the target distribution, \(p(s_{i}|L_{T} = i\)), and parameters of the distractor distribution, \(p(s_{i}|L_{T} \neq i\)), for each location \(i = 1\ldots N\) in the stimuli matrix as well as the target location, \(L_{T}\). These parameters were then used to generate the stimuli \(s_{i}\) at each location.

Neither the task parameters nor the stimuli are known to the Bayesian observer. Instead, at each moment in time \emph{t} within a trial, the observer obtains sensory observations at each location, \(x_{i,t}\). These observations are not identical to the stimuli because of the presence of sensory noise, \(p\left( x_{i,t} \mid s_{i} \right)\). That is, a given stimulus might result in different sensory responses, and, conversely, a given sensory observation might correspond to different stimuli. We assume that the observations are distributed independently at each location and at each moment in time.

To make an optimal decision in a particular task, the observer needs to know the relationship between the sensory observations and the task-relevant quantities. For the visual search task used in our study, we assumed that observers compare for each location the probability that the sensory observations are caused by a target present at that location, \(p\left( L_{T} = i \mid \mathbf{x}_{i} \right)\) where \(\mathbf{x}_{i} = \{ x_{i,1},x_{i,2},\ldots,x_{i,t = K}\}\) are the samples obtained for location \emph{i} up until the time \emph{K}, against the probability that they are caused by a distractor, \(p\left( L_{T} \neq i \mid \mathbf{x}_{i} \right)\):
\begin{equation} \label{eq:9}
d_{i} = \frac{p\left( L_{T} = i \mid \mathbf{x}_{i} \right)}{p\left( L_{T} \neq i \mid \mathbf{x}_{i} \right)}
\end{equation}

The observer then decides that a given item is a target as soon as the decision variable at a given location reaches a certain threshold \(B\). Although this decision rule is not fully optimal, because the observer makes a decision for each item individually, it greatly reduces the task complexity, and we believe that it allows for a more realistic model (the simulations based on a more complex but more optimal model are described in the supplement and lead to identical conclusions).

The observer can compute the probability of hypotheses \(L_{T} = i\) and \(L_{T} \neq i\) given the sensory data using the Bayes rule:
\begin{equation} \label{eq:10}
p\left( L_{T} = i \mid \mathbf{x}_{i} \right) = \frac{p\left( \mathbf{x}_{i} \mid L_{T} = i \right)p\left( L_{T} = i \right)}{p\left( \mathbf{x}_{i} \right)}
\end{equation}
In words, the probability of a hypothesis that a target is at the given location, \(L_{T} = i\), for a set of sensory observations \(\mathbf{x}_{i}\) is equal to the likelihood of the data given this hypothesis multiplied by a prior probability for this hypothesis \(p\left( L_{T} = i \right)\) and divided by the probability of the observations \(p\left( \mathbf{x}_{i} \right)\).

Assuming that the prior \(p\left( L_{T} = i \right) = \frac{1}{N} = 1 - p\left( L_{T} \neq i \right)\) is the same for all locations, the decision variable can then be rewritten in log-space as the difference in the log-likelihoods in favor of the two hypotheses:
\begin{equation} \label{eq:11}
\log d_{i} = \log\left( \frac{1}{N - 1} \right) + \sum_{t = 1}^{K}{\log\left( p\left( x_{i,t} \mid L_{T} = i \right) \right)} - \sum_{t = 1}^{K}{\log{ p\left( x_{i,t} \mid L_{T} \neq i \right) }} 
\end{equation}

What are the probabilities of sensory observations under each hypothesis, \(p\left( x_{i,t} \mid L_{T} = i \right)\) and \(p\left( x_{i,t} \mid L_{T} \neq i \right)\)? To compute them, the observer needs to take into account how the stimuli are distributed under each hypothesis and how the sensory noise is distributed for each stimulus. We assume that the sensory noise distribution is known to the observer through long-time exposure to the visual environment (that is, the observer knows \(p\left( x_{i,t} \mid s_{i} \right)\)).

However, to determine how probable it is that sensory observations correspond to the search target, the observer must also know what defines targets and distractors. The experimenter knows that only certain orientations describe a target, but the observer is not omniscient and does not know the true distributions of target and distractor stimuli, approximating them instead as \(p^{*}\left( s_{i} \mid L_{T} = i \right)\) and \(p^{*}\left( s_{i} \mid L_{T} \neq i \right)\). Then the probability of sensory observations under each hypothesis can be computed as:
\begin{equation} \label{eq:12}
p\left( x_{i,t} \mid L_{T} \neq i \right) = \int_{}^{}{p\left( x_{i,t} \mid s_{i} \right)p^{*}\left( s_{i} \mid L_{T} \neq i \right)\,ds_{i}}
\end{equation}

The probability distributions \(p^{*}\left( s_{i} \mid L_{T} = i \right)\) and \(p^{*}\left( s_{i} \mid L_{T} \neq i \right)\) correspond to the observer's approximate representation of target and distractor distributions. Notably, each of them can be further separated into the representation based on the previous trials and the one based on the current trial:
\begin{equation} \label{eq:13}
p^{*}\left( s_{i} \mid L_{T} \neq i \right) \equiv p^{*}\left( s_{i} \mid L_{T} \neq i,\bm{\uptheta} \right) = p^{*}\left( s_{i} \mid L_{T} \neq i,\bm{\uptheta}_{prev} \right)p^{*}\left( s_{i} \mid L_{T} \neq i,\bm{\uptheta}_{curr} \right)
\end{equation}
with \(\bm{\uptheta}=\left\{ \bm{\uptheta}_{prev},\bm{\uptheta}_{curr} \right\}\) corresponding to the independent latent variables describing the parameters of the previous and the current trial by the observer (similar equations related to targets are omitted for brevity). In our experiments, by design, the parameters of the current trial are controlled with respect to the current stimuli (i.e., the distractors on the current test trial are drawn from a distribution with a mean from 60° to 120° off the current test target). Hence, only \(p^{*}\left( s_{i} \mid L_{T} \neq i,\bm{\uptheta}_{prev} \right)\) matters for relative changes in response times.

In our analyses, we wanted to reconstruct the representation of distractor stimuli using the response times for different test targets. Because the decision time is proportional to the number of samples when the sampling frequency is constant, we aimed to relate the number of samples \emph{K} to an observer's approximate representation of distractors based on the previous trials \(p^{*}\left( s_{i} \mid L_{T} \neq i,\bm{\uptheta}_{prev} \right)\).

Assuming that the sensory observations are obtained with high frequency, we can approximate the total evidence in favor of a given hypothesis:
\begin{equation} \label{eq:14}
\sum_{t = 1}^{K}{\log{ p\left( x_{i,t} \mid L_{T} \neq i \right) }} \approx K\left( E\left\lbrack \log{p\left( x_{i,t} \mid L_{T} \neq i \right)}  \right\rbrack \right)
\end{equation}

We expect the sensory noise to be low compared to the noise in the target and distractor representations. Then, the following approximation is valid:
\begin{equation} \label{eq:15}
E\left\lbrack \log{ p\left( x_{i,t} \mid L_{T} \neq i \right) } \right\rbrack \propto \log{ p^{*}\left( s_{i} \mid L_{T} \neq i \right) } + C
\end{equation}
where \emph{C} is a constant. Similar derivations can be used for the total evidence for the alternative hypothesis \(p\left( x_{i,t} \mid L_{T} = i \right)\).

Then, given that a decision is made when \(\log d_{i} = \log B\):
\begin{equation} \label{eq:16}
K = \frac{\log B - \log\left( \frac{1}{N - 1} \right)}{E\left\lbrack \log{ p\left( x_{i,t} \mid L_{T} = i \right) } \right\rbrack - E\left\lbrack \log{ p\left( x_{i,t} \mid L_{T} \neq i \right) } \right\rbrack}
\end{equation}

Given that the target and distractor parameters are independently manipulated in the experiment, \(E\left\lbrack \log\left( p\left( x_{i,t} \mid L_{T} = i \right) \right) \right\rbrack\) can be treated as a constant. Similarly, \(p^{*}\left( s_{i} \mid L_{T} = i,\bm{\uptheta}_{curr} \right)\) would be constant as discussed above. Given that \(RT \propto K\), we can then approximate is as follows:
\begin{equation} \label{eq:17}
RT \approx \frac{C_{1}}{C_{0} - \log{p^{*}\left( s_{i} \mid L_{T} \neq i,\bm{\uptheta}_{prev} \right)}}
\end{equation}
and
\begin{equation} \label{eq:18}
\log{p^{*}\left( s_{i} \mid L_{T} \neq i,\bm{\uptheta}_{prev} \right)} = C_{0} - C_{1}\frac{1}{RT}
\end{equation}
where \(C_{0}\) and \(C_{1}\) are constants. In words, there is an inverse linear relationship between the likelihood that a given stimulus is a distractor (in log-space) and the response times. When this likelihood increases, response times decrease.

We highlight that this model provides an important insight, namely, that observers' representations are monotonically related to response times. Hence, even though \(C_{0}\) and \(C_{1}\) are unknown, the relationship between the moments (mean, standard deviation, and skewness) of observers' representations reconstructed from RT and the true representations would hold under any other monotonic transformation (for example, RTs are log-transformed and the baseline RTs are subtracted as we do in our analyses).

\section*{Acknowledgements}
We are grateful to Alena Begler for the help with data collection and to James Cooke for his invaluable comments on the models included in the manuscript.

\hypertarget{data-availability}{\section*{Data availability}\label{data-availability}}

The data and scripts used for the data analysis in this paper are available from \url{https://osf.io/5pfyn/}.

\section*{Conflict of interest}
The authors declare no conflict of interest. 

\printbibliography

\renewcommand{\theequation}{S.\arabic{equation}}
\setcounter{equation}{0}
\clearpage

\hypertarget{supplement-1.-bayesian-observer-model-combining-information-across-locations.}{%
\section*{Supplement 1. Bayesian observer model combining information across locations.}\label{supplement-1.-bayesian-observer-model-combining-information-across-locations.}}

The model reported in the main text presents a simplified version of the decision-making process assuming that stimuli at each location are analyzed separately. We believe that such a model might be more realistic as it greatly simplifies the computations that observers have to make. However, for the sake of completeness, here we briefly describe a more complex conditionally-optimal memory-guided Bayesian observer model. We refer to this model as conditionally optimal for two reasons. First, a memory-guided observer is by definition not fully optimal in our task, where the test trial parameters are unrelated to the previous learning trials. However, given that the task parameters repeat throughout learning trials, using the information from the previous trials might be beneficial when the observer does not know that the trial parameters have changed. Secondly, we assume that the observer's learning or memory about the stimuli features might not be ideal, hence they use the approximations of feature distributions. We show that under this more complex and more optimal model, the predictions with respect to the monotonic relationship between the response times and expected distractor probabilities stay the same.

\subsection*{Task structure}
Participants have to locate a target among a set of distractors and indicate if it is in the top or in the lower part of the stimuli matrix. The experimenter sets the task parameters, namely, the target distribution, \(p(s_{i}|L_{T} = i\)), and the distractor distribution, \(p(s_{i}|L_{T} \neq i\)), for each location \(i = 1\ldots N\) in the stimuli matrix (with top half having indices from 1 to \(N/2\) and the bottom half from \(\frac{N}{2} + 1\) to \emph{N}) as well as the target location (\(L_{T}\)), to generate the stimuli (\(s_{i}\)) at each location. Here, \(L_{T} = i\) and \(L_{T} \neq i\) indicate that the target is or is not at location \emph{i}, or in other words, that the target location is or is not \emph{i}, respectively.

\subsection*{Ideal observer model}
At each moment in time \(t = 1\ldots K\) (with \emph{K} as the decision moment) and at each location \emph{i}, the observer obtains sensory observations \(x_{i,t}\) corrupted by the presence of sensory noise:
\begin{equation} \label{eq:svm}
p\left( x_{i,t} \mid s_{i} \right) = f_{VM}(x_{i};s_{i},\kappa_{s})
\end{equation}
where \(f_{VM}\) is a von Mises distribution density with concentration parameter \(\kappa_{s}\) quantifying the amount of noise. We assume that the observations are distributed independently at each location and at each moment in time:
\begin{equation} \label{eq:s1}
p\left( \mathbf{X} \mid \mathbf{s} \right) = \prod_{i = 1}^{N}{p\left( \mathbf{x}_{\mathbf{i}} \mid s_{i} \right)} = \prod_{i = 1}^{N}{\prod_{t = 1}^{K}{p\left( x_{i,t} \mid s_{i} \right)}}
\end{equation}

To make an optimal decision in a particular task, the observer needs to compare the probability that a target is located in the upper half of the stimuli matrix with a probability that it is located in the lower half:
\begin{equation} \label{eq:s2}
d = \frac{p\left( C = 1 \mid \mathbf{X} \right)}{p\left( C = 2 \mid \mathbf{X} \right)}
\end{equation}
where \(C = 1\) and \(C = 2\) correspond to the two hypotheses about the target location. After applying the log transformation, the decision variable can be expressed as a difference in the amount of evidence for the two hypotheses:
\begin{equation} \label{eq:s3}
\log d = \log{p\left( C = 1 \mid \mathbf{X} \right)} - \log{p\left( C = 2 \mid \mathbf{X} \right)}
\end{equation}

The decision time assuming a certain threshold \emph{B} can then be found as a time \emph{K} when the decision variable reaches the threshold. The average decision time can be found by estimating when the expectation of \(\log d\) becomes equal to \(\log B\):
\begin{equation} \label{eq:s4}
K = \frac{\log B}{E\left\lbrack \log{p\left( C = 1 \mid \mathbf{X} \right)} \right\rbrack - E\left\lbrack \log{p\left( C = 2 \mid \mathbf{X} \right)} \right\rbrack}
\end{equation}

The probabilities for each hypothesis \(C = 1\) and \(C = 2\) can be found using the Bayes rule. For example, for \(C = 1\):
\begin{equation} \label{eq:s5}
p\left( C = 1 \mid \mathbf{X} \right)=\frac{p\left( \mathbf{X} \mid C = 1 \right)p(C = 1)}{p\left( \mathbf{X} \right)}
\end{equation}

Because the observer does not know what stimuli are presented and only knows the sensory observations, the likelihood \(p\left( \mathbf{x} \mid C = 1 \right)\) needs to be computed by averaging (marginalizing) over the unknown stimuli values:
\begin{equation} \label{eq:s6}
p\left( \mathbf{X} \mid C = 1 \right) = \int_{}^{}{p\left( \mathbf{X} \mid \mathbf{s} \right)p\left( \mathbf{s} \mid C = 1 \right)d\mathbf{s}}
\end{equation}

Because the target can only be present at one location, the likelihood \(p\left( \mathbf{x} \mid C = 1 \right)\) is computed by summing over the possibilities of finding a target at each particular location:

\begin{equation} \label{eq:s7}
p\left( \mathbf{X} \mid C = 1 \right) = \sum_{i = 1}^{\frac{N}{2}}{\int_{}^{}{p\left( \mathbf{X} \mid \mathbf{s} \right)p^{*}\left( \mathbf{s} \mid L_{T} = i,\bm{\uptheta} \right)d\mathbf{s}}}
\end{equation}
where similarly to the main text, we use an asterisk to denote probability distributions as approximated by the observer through a set of parameters related to previous and current trials \(\bm{\uptheta}= \{\bm{\uptheta}_{prev},\bm{\uptheta}_{curr}\}\). That is, we assume that the observer is unaware of the true distributions \(p(s_{i}|L_{T} = i\)) and \(p(s_{i}|L_{T} \neq i\)) and approximates them instead using the information available. Note that the sum is done separately for each half of the stimuli matrix, hence $\frac{N}{2}$ is used in Eq. \ref{eq:s7}. 

If a target is at location \emph{i,} it cannot be anywhere else. Hence:
\begin{equation} \label{eq:s8}
p^{*}\left( \mathbf{s} \mid L_{T} = i,\bm{\uptheta} \right) = p^{*}\left( s_{i} \mid L_{T} = i,\bm{\uptheta} \right)\prod_{j \neq i}^{N}{p^{*}\left( s_{j} \mid L_{T} \neq j,\bm{\uptheta} \right)}
\end{equation}

Using Eq. \ref{eq:s8}, it can be further shown that:
\begin{equation} \label{eq:s9}
\int_{}^{}{p\left( \mathbf{X} \mid \mathbf{s} \right)p^{*}\left( \mathbf{s} \mid L_{T} = i,\bm{\uptheta} \right)d\mathbf{s}} = \left\lbrack \prod_{j}^{N}{\int_{}^{}{p\left( \mathbf{x}_{j} \mid s_{j} \right)p^{*}\left( s_{j} \mid L_{T} \neq j,\bm{\uptheta} \right)ds_{j}}} \right\rbrack\frac{\int_{}^{}{p\left( \mathbf{x}_{i} \mid s_{i} \right)p^{*}\left( s_{i} \mid L_{T} = i,\bm{\uptheta} \right)ds_{i}}}{\int_{}^{}{p\left( \mathbf{x}_{i} \mid s_{i} \right)p^{*}\left( s_{i} \mid L_{T} \neq i,\bm{\uptheta} \right)ds_{i}}}
\end{equation}
Note that the product in the square brackets is the same for all locations, and the remaining part of the equation is a ratio of the probability that the measurements at a given location are from the target against the probability that they are from the distractor, similarly to the model described in the main text.

The probability that a given stimulus is a target (or a distractor) depends on both the previous and the current trial:
\begin{equation} \label{eq:s10}
p^{*}\left( s_{i} \mid L_{T} = i,\bm{\uptheta} \right) = p^{*}\left( s_{i} \mid L_{T} = i,\bm{\uptheta}_{prev} \right)p^{*}\left( s_{i} \mid L_{T} = i,\bm{\uptheta}_{curr} \right)
\end{equation}

For each location and each location-specific hypothesis \(L_{T} = i\) and \(L_{T} \neq i\), the current trial parameters need to be computed separately because of the nature of the odd-one-out task. A target is defined as the item most different from the distractors. For simplicity, we assumed that observers use the following circular normal approximation for the distractors at the current trial based on the sensory observations:
\begin{equation} \label{eq:s11}
p^{*}\left( s_{i} \mid L_{T} \neq i,\bm{\uptheta}_{curr} \right) = f_{VM}\left( s_{i};{\hat{\mu}}_{j \neq i},{\hat{\kappa}}_{j \neq i} \right)
\end{equation}
In words, when the observer needs to estimate, how likely it is that the stimulus at location \emph{i} is a distractor, the observer approximates the distribution of stimuli as a von Mises (circular normal) distribution based on the sensory observations from other locations.

The observer might use the knowledge that the target distribution in the task design is on average 90° away from the mean of distractors. We again assume a von Mises approximation:
\begin{equation} \label{eq:s12}
p^{*}\left( s_{i} \mid L_{T} = i,\bm{\uptheta}_{curr} \right) = f_{VM}\left( s_{i};{\hat{\mu}}_{j \neq i} + 90{^\circ},\kappa_{T} \right)
\end{equation}
where \(\kappa_{T}\) is the expected precision of the target distribution. In contrast to the distractor distribution precision that could be guessed based on the samples on the current trial (\({\hat{\kappa}}_{j \neq i}\)), the target distribution precision cannot be estimated on a single trial (there is only one target stimulus in a given trial) and has to be based on the other sources of information (e.g., learning throughout the experiment).

Given that the measurement noise is independent across locations, the likelihood of the hypothesis \emph{C} = 1 can be further expressed as:
\begin{equation} \label{eq:s13}
p\left( \mathbf{X} \mid C = 1 \right) = \left\lbrack \prod_{j = 1}^{N}{\int_{}^{}{\left( \mathbf{x}_{j} \mid s_{j} \right)p^{*}\left( s_{j} \mid L_{T} \neq j,\bm{\uptheta} \right)ds_{j}}} \right\rbrack\sum_{i = 1}^{\frac{N}{2}}\frac{\int_{}^{}{p\left( \mathbf{x}_{i} \mid s_{i} \right)p^{*}\left( s_{i} \mid L_{T} = i,\bm{\uptheta} \right)ds_{i}}}{\int_{}^{}{p\left( \mathbf{x}_{i} \mid s_{i} \right)p^{*}\left( s_{i} \mid L_{T} \neq i,\bm{\uptheta} \right)ds_{i}}}
\end{equation}

Then, assuming that the prior probability of each decision alternative is the same, the decision variable can be expressed in log-space as:
\begin{equation} \label{eq:s14}
\log d = \log\left( \sum_{i = 1}^{\frac{N}{2}}\frac{\int_{}^{}{p\left( \mathbf{x}_{i} \mid s_{i} \right)p^{*}\left( s_{i} \mid L_{T} = i,\bm{\uptheta} \right)ds_{i}}}{\int_{}^{}{p\left( \mathbf{x}_{i} \mid s_{i} \right)p^{*}\left( s_{i} \mid L_{T} \neq i,\bm{\uptheta} \right)ds_{i}}} \right) - \log\left( \sum_{i = \frac{N}{2}+1}^{N}\frac{\int_{}^{}{p\left( \mathbf{x}_{i} \mid s_{i} \right)p^{*}\left( s_{i} \mid L_{T} = i,\bm{\uptheta} \right)ds_{i}}}{\int_{}^{}{p\left( \mathbf{x}_{i} \mid s_{i} \right)p^{*}\left( s_{i} \mid L_{T} \neq i,\bm{\uptheta} \right)ds_{i}}} \right)
\end{equation}

The decision time assuming a certain threshold \emph{B} can then be found as a time \emph{K} when the decision variable reaches the threshold.

\subsection*{Simulations} 

To estimate the behavior of the observer using this model, we simulated the decision-making process and estimated the mean response times while varying the properties of the distractor representation \(p^{*}\left( s_{i} \mid L_{T} \neq i,\bm{\uptheta}_{prev} \right)\). The task parameters were based on the actual experiment design. We used 36 stimuli for each trial with one stimulus being the test target (\(s_{L_{T}}\)) and the rest being the distractors. The distractors on each simulated trial were distributed as \(p\left( s_{i} \mid L_{T} \neq i \right) = f_{VM}\left( s_{i};\mu_{D},\kappa_{D} \right)\) where \(\mu_{D}\sim U\left( s_{L_{T}} + 60{^\circ};s_{L_{T}} + 120{^\circ} \right)\) (that is, the mean of distractors is set to 60° to 120° away from the test stimulus) and \(\kappa_{D} = 8.7\) (approximately equivalent to the standard deviation of 10° in orientation space). The sensory observations were assumed to be noisy (\(\kappa_{s} = 2\), approximately equivalent to the standard deviation of 24° in orientation space; note that this is the noise level for samples collected at each moment in time). The observers' target representation was assumed to be linked with to the distractor representation as \(p^{*}\left( s_{i} \mid L_{T} = i,\bm{\uptheta}_{prev} \right) = f_{VM}\left( s_{i};\mu_{D_{prev}},\kappa_{T} \right)\) with \(\kappa_{T} = 3.35\) (based on a normal approximation to a uniform target distribution with 60° range used in the experiments). The same \(\kappa_{T}\) was used for target-related computations based on the current trial data (Eq. \ref{eq:s12}). The decision threshold was set to \(\log B = 4.60\) assuming a 1\% probability of error if the observer assumptions are correct. For each test target from 1° to 180° in half-degree steps, we simulated 56 trials for each combination of distractor representation parameters.

We ran simulations for the wrapped skewed normal distribution with the mean varied from -60° to 60° in 20° steps, while the standard deviation varied from 20° to 60° in 10° steps, and skew varied from -10 to 10 in steps of 2. The results of the simulations (Figure \ref{fig:sim_adequate_IO}) confirmed the findings obtained with a simplified model: the means are recovered precisely while for standard deviation and skewness the monotonic relationship holds.

\renewcommand{\thefigure}{S\arabic{figure}} 
\setcounter{figure}{0}
\clearpage
\begin{figure}[t]
\centering
\includegraphics[width=\textwidth,height=\textheight,keepaspectratio]{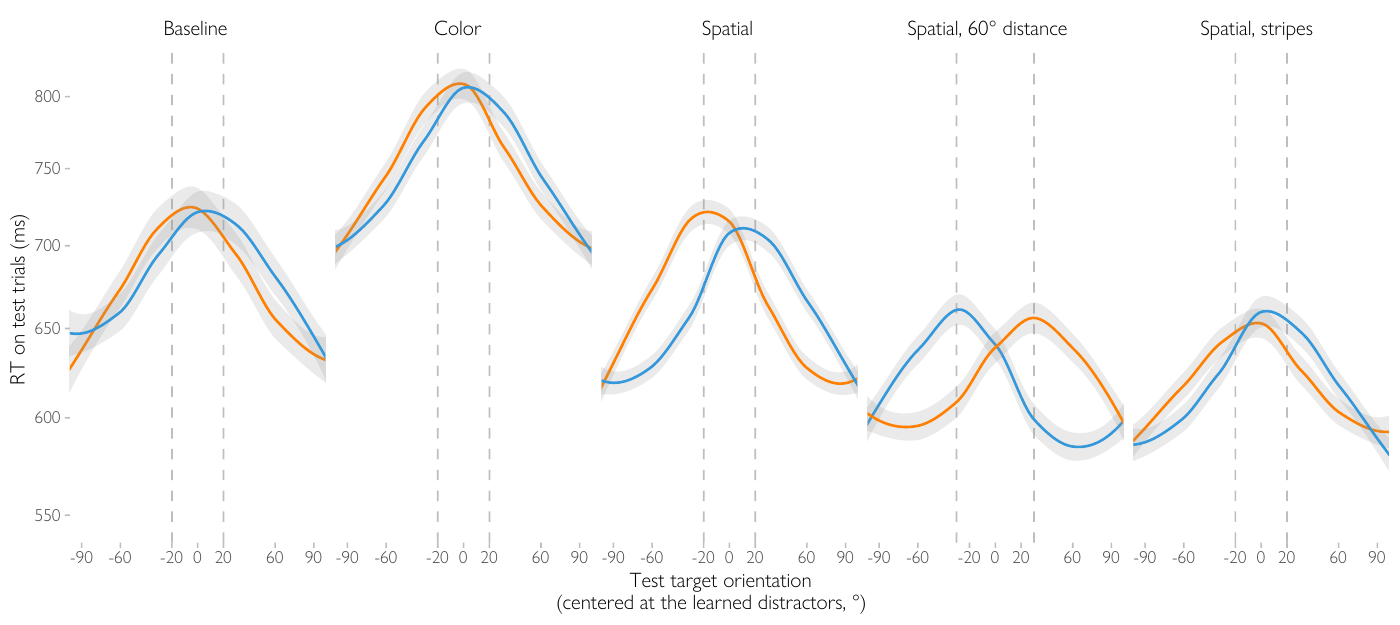}
\caption{Raw response times for test trials in different conditions. The curves shows the average RT with 95\% CI shown with shading. Different colors indicate RT for test trials when a test target matches (in location or color) the clockwise- and counterclockwise-shifted parts of the learned distractor distribution. Dashed lines show the mean orientations for the corresponding distribution parts. The average was estimated with locally-weighted regression (LOESS, an extension of the moving window approach) that accounted for circularity of the orientation space by padding the data at each end of the [-90,90] range with 1/6\textsuperscript{th} of the data from the other end.}
\label{fig:suppl_rawRT}
\end{figure}
\null
\vfill

\clearpage
\begin{figure}[t]
\centering
\includegraphics[width=\textwidth,height=\textheight,keepaspectratio]{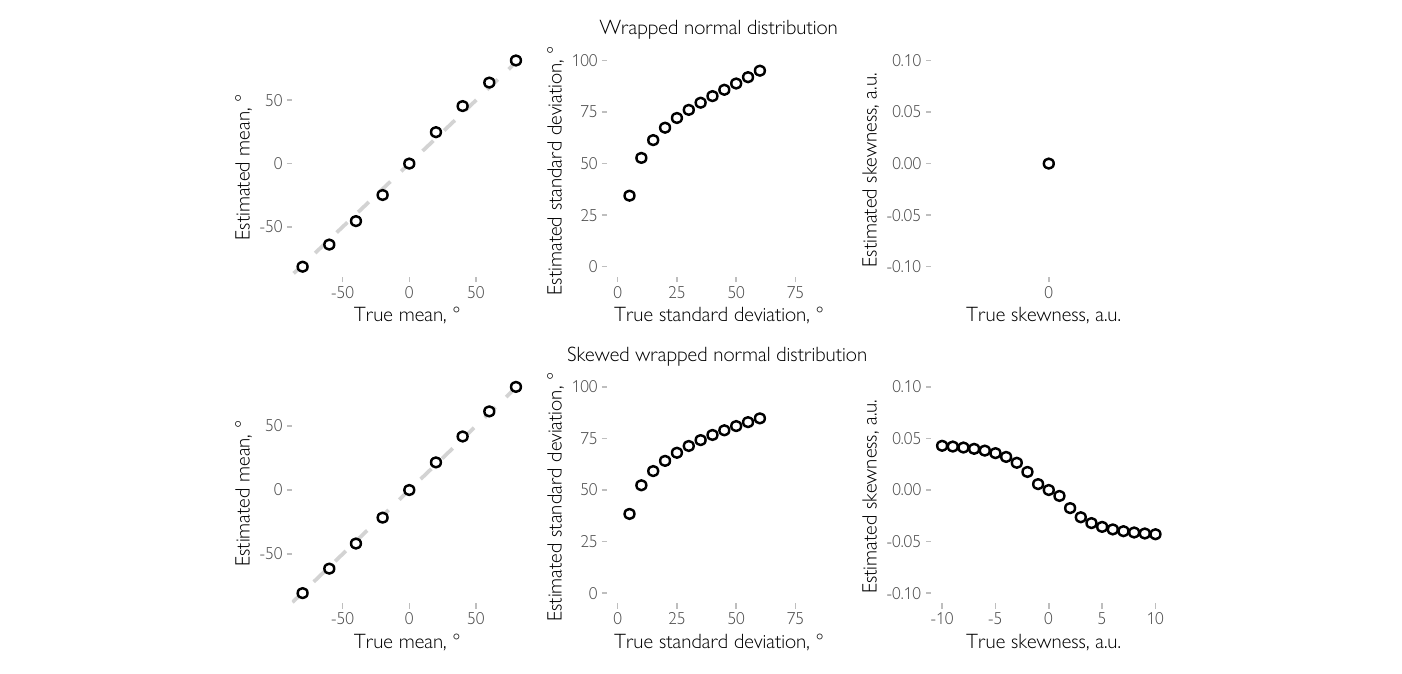}
\caption{Simulated parameters under the simplified Bayesian observer model. We simulated the response times under the assumptions of the simplified Bayesian observer model described in the main text and applied the same approach as used for the real data to see if the assumed monotonic relationship between the true parameters and the recovered parameters holds. Firstly, we used a simple wrapped normal (top) with means varying from -80° to 80° in 20° steps and standard deviation from 5° to 60° in 5° steps. For each parameter combination the RT were computed using Eq. 16. We then estimated the parameters of the recovered distribution. As is evident from the plots, the mean estimates were identical to the true mean while the standard deviation was overestimated but the overall monotonic relationship held. The skewness estimate was at zero as expected for the symmetric wrapped normal distribution. Secondly, we simulated the data using the skewed normal distribution \citep{pewsey2004Large} with means again varying from -80° to 80° in 20° steps, scale parameter varying from 5° to 60° in 5° steps, and skewness parameter varying from -10 to 10 in steps of 1. For the means and standard deviations, the conclusions were the same as for the wrapped normal distribution. Similarly, skewness estimates followed monotonically the changes in the true skewness parameter (note that the sign of the estimated circular skewness is the opposite of the skewness parameter of the skewed wrapped normal distribution because of how it is defined, see \citealp{pewsey2004Large}). In sum, the mean estimates match the true means, and the standard deviation and skewness estimates monotonically depend on the true standard deviation and the skewness parameters.}
\label{fig:sim_simpl_IO}
\end{figure}
\null
\vfill

\clearpage
\begin{figure}[t]
\centering
\includegraphics[width=\textwidth,height=\textheight,keepaspectratio]{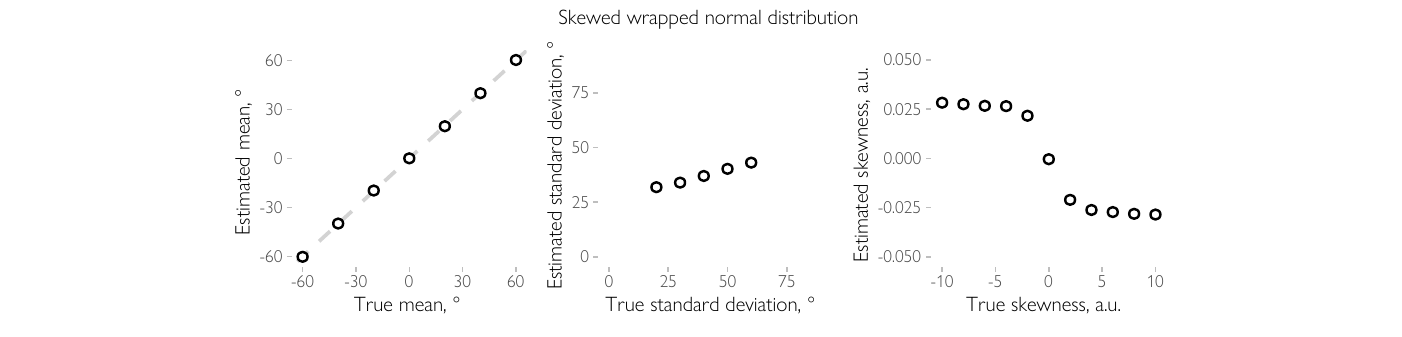}
\caption{Simulated parameters under the more optimal Bayesian observer model. We simulated the response times under the assumptions of the more complex Bayesian observer model described in the Supplement applied the same approach as used for the real data to see if the assumed monotonic relationship between the true parameters and the recovered parameters holds. The results were similar to the simulations with the simplified model (Figure \ref{fig:sim_simpl_IO}). The mean estimates were identical to the true mean, while for the standard deviation and skewness the monotonic relation holds (note that the sign of the estimated circular skewness is the opposite of the skewness parameter of the skewed wrapped normal distribution because of how it is defined, see  \citealp{pewsey2004Large}). In sum, the mean estimates match the true means, and the standard deviation and skewness estimates monotonically depend on the true standard deviation and the skewness parameters.}
\label{fig:sim_adequate_IO}
\end{figure}
\null
\vfill

\end{document}